\documentclass[12pt]{article}

\usepackage[a4paper,left=30mm,right=20mm,top=20mm,bottom=20mm]{geometry}
\usepackage{graphics}
\usepackage{amsmath}
\usepackage{epsfig}
\usepackage{cite}
\usepackage{xcolor}
\newcommand{\z}{&&\hspace*{-1cm}}

\newcommand{\ep}{\varepsilon}

\newcommand{\bea}{\begin{eqnarray}}
\newcommand{\eea}{\end{eqnarray}}
\newcommand{\be}{\begin{equation}}

\newcommand{\ee}{\end{equation}}
\newcommand{\MSbar}{\overline{\rm MS}}

\title{Parametrizations of collinear and $k_T$-dependent
  parton densities in a proton
}

\author{N.A.~Abdulov$^{1}$, A.V.~Kotikov$^{2}$, A.V.~Lipatov$^{1,2}$}

\begin{document}

\maketitle

\begin{center}
%{\it $^{1}$Institute of Modern Physics, Lanzhou 730000, China}\\
{\it $^{1}$Skobeltsyn Institute of Nuclear Physics, Lomonosov Moscow State University, 119991, Moscow, Russia}\\
{\it $^{2}$Joint Institute for Nuclear Research, 141980, Dubna, Moscow region, Russia}
%\\
%{\it $^{4}$ School of Physics and Astronomy, Sun Yat-sen University, Zhuhai 519082, China}

\end{center}

\vspace{0.5cm}

\begin{center}

{\bf Abstract }
       
\end{center}

\indent
A new type of parametrization for parton distribution 
functions in a proton,
based on their $Q^2$-evolution at large and small $x$ values, is
constructed. 
In our analysis, the valence and nonsinglet parts obey 
the Gross-Llewellyn-Smith and Gottfried sum rules, respectively. For the singlet 
quark and gluon densities momentum conservation 
is taken into account. 
Then, using the Kimber-Martin-Ryskin prescription,
we extend the consideration
to Transverse Momentum Dependent (TMD, or unintegrated) 
gluon and quark distributions in a proton,
which currently plays an important role in a 
number of phenomenological applications.
The analytical expressions for the latter, valid for both low and large $x$,
are derived for the first time.

%in the parametrizations.

\vspace{1.0cm}

\noindent{\it Keywords:}
%small $x$, large $x$, QCD evolution
%, TMD parton densities, high-energy factorization.
QCD evolution, parton density functions in a proton, Kimber-Martin-Ryskin approach

\newpage

\section{Introduction} \indent

The parton (quark and gluon) distribution functions (PDFs) in a proton 
are necessary part of any theoretical study performed 
within the Quantum Chromodynamics (QCD).
They encode the information on the non-perturbative 
structure of a proton and directly related to the calculated
cross sections (or other observables)
via certain QCD factorization theorem.
The QCD evolution leads to their essential dependence on the 
probing scale $Q^2$, which can be described by the 
Dokshitzer-Gribov-Lipatov-Altarelli-Parisi (DGLAP) equations\cite{DGLAP}.
Usually, the latter is solved numerically with leading order (LO),
next-to-leading (NLO) or even next-to-next-to-leading (NNLO) accuracy,
where number of corresponding phenomenological parameters of initial 
parton distributions are fitted at HERA, LHC and fixed target experiments
at various  $(x,Q^2)$ ranges\footnote{Several recent PDF fits, 
as well as the references to the previous
studies, can be found\cite{NNLOfits,PKK,BKKP}.}. 
Large uncertainties for many processes at the LHC
originate, of course, mainly from our restricted knowledge 
of the parton distributions (see, for example,\cite{Forte} and references therein).
Thus, studying the proton PDFs from both theoretical and 
experimental points of view is an important and urgent task.

In the present paper we continue with
the idea\cite{Illarionov:2010gy}
to present more information on the PDFs from the theoretical side. The approach consists of
the two basic steps.
First, we find asymptotics of solutions of
the DGLAP equations for the parton densities at small and large values of the
Bjorken variable $x$. Second,
we combine these two solutions and then interpolate between them to obtain 
the analytical expressions for PDFs
over the full range of $x$.

In a sense, this is not a very new idea. A similar approach
has been proposed\cite{LoYn,Yndu} % (see also references therein) 
about of $50$ years ago. However, in the present paper the 
parametrizations are constructed in a rather
different way. In particular, following\cite{Illarionov:2010gy},
they include important subasymptotic terms which
are fixed exactly by the momentum conservation and also by 
the  Gross-Llewellyn-Smith and Gottfried sum rules (see\cite{GrLl} and \cite{Gottfried:1967kk}, respectively).
Such calculations will be performed for the 
first time
providing the community with new type of 
parametrization of gluon and 
quark densities in a proton valid at low and large $x$.
%Then, using the Kimber-Martin-Ryskin formalism\cite{?}, 
Moreover,
we extend our consideration\cite{Kotikov:2019kci,Kotikov:2021yzb} and derive 
analytical expressions for 
Transverse Momentum Dependent (TMD)
parton distributions
using the Kimber-Martin-Ryskin (KMR) framework\cite{21,22}.
These quantities are known to be a very suitable tool to investigate a
less inclusive processes which proceed at high energies with
large momentum transfer and/or containing multiple hard scales (see, for example, review\cite{TMDReview} 
and references therein). 
Our main motivation is that up to now there is no
analytical expressions for gluon and especially quark TMDs (both sea and valence)
valid in a wide $x$ region\footnote{In our previous study\cite{Kotikov:2019kci,Kotikov:2021yzb}, 
only small $x$ limit have been considered and 
phenomenological model for large $x$ region have been applied.}.

The analysis of the present paper is limited to the LO in the 
perturbation theory, which 
%that
is reasonable\cite{Sherstnev}
%important \cite{Sherstnev} because
for those processes
at the LHC for which the NLO corrections are not known at present.
Moreover, most of phenomenological applications 
involving TMDs are currently performed at the LO also (see, 
for example,\cite{kt-app1, kt-app2, Kotikov:2021yzb, kt-app4, kt-app5} and references therein).
%This difference is much more significant\cite{Kotikov2007} than the
%differences between PDFs extracted by various groups.
%It is clear that the LO parton densities should be used with the
%LO matrix elements. 
On the other hand, the consideration of PDFs at LO is the 
necessary first step in studying PDFs and TMDs
at higher orders. These higher order
corrections can be treated like those\cite{Yndu,Neubert,Buras}.

The outline of our paper is following. In Section 2 we describe our theoretical
input. Sections~3 and 4 contain low $x$ and large $x$ PDF asymptotics.
Parametrizations of parton densities, their properties and numerical results for PDFs are given 
in Section~5. Section~6 is devoted to TMD parton densities in the Kimber-Martin-Ryskin framework.
Section~7 contains our conclusions. Most complicated calculations are presented in Appendices.
%A,B and C.}

\section{ Theoretical input } \indent

%Here we briefly touch on certain aspects of the theoretical part of our
%analysis. For a bit detailed account see, for example, \cite{Kotikov2007}.

%\iffalse
In this section we briefly present the theoretical part of our analysis.
The reader is referred to \cite{Kotikov2007} for more details.

The deep-inelastic scattering (DIS)
%cross-section
$l + N \to l^\prime + X$, where
$l$ and $N$ are %denote
%mark i
the incoming lepton and nucleon, and $l^\prime$ is the outgoing lepton,
in one of the basic processes for studying of the nucleon structure. The DIS cross-section
can be split to the lepton
$L^{\mu \nu}$ and hadron $F^{\mu \nu}$ parts
\begin{equation}
d\sigma \sim L^{\mu \nu} F^{\mu \nu} \, .
\label{S1.1}
\end{equation}

%Because photon is
The lepton part $L^{\mu \nu}$
%can be
is evaluated exactly, while the hadron one,
$F^{\mu \nu}$, can be presented in the following form
\begin{eqnarray}
F^{\mu \nu} &=& \left(-g^{\mu \nu}+ \frac{q^{\mu}q^{\nu}}{q^2}\right) \, F_1(x,Q^2)
%\nonumber \\&+&
+\left(p^{\mu}+ \frac{(pq)}{q^2}q^{\mu}\right) \left(p^{\nu}+
\frac{(pq)}{q^2}q^{\nu}\right) \, F_2(x,Q^2)
\nonumber \\
&+& i\varepsilon_{\mu \nu \alpha \beta}p^{\alpha}p^{\beta} \frac{x}{q^2}
\, F_3(x,Q^2) + \dots \,,
\label{S1.2}
\end{eqnarray}
where the symbol $\dots$ stands for those parts which depend on 
the nucleon spin.
The functions
$F_k(x,Q^2)$ with $k=1$, $2$ and $3$ are the DIS structure functions (SFs) and
$q$ and $p$ are the photon and parton
%hadron (parton)
momenta.
%\fi
Moreover, the two variables
%the values
\begin{equation}
Q^2=-q^2 > 0 ~,~~ x=\frac{Q^2}{2(pq)}
\label{S1.var}
\end{equation}
determine the basic properties of the DIS process. Here, $Q^2$ is the ``mass''
of the virtual photon and/or $Z/W$ boson, and the Bjorken variable $x$
($0<x<1$) is the part of the hadron momentum carried by the scattering parton (quark or gluon).

%It is very useful
%{\bf 1.}~~

\subsection{Mellin transform} \indent

%\iffalse
%Now we move to Mellin transform, which
The Mellin transform
%is very important for DIS scattering.
diagonalizes the $Q^2$ evolution of the parton densities.
%(see Eq.~(\ref{S2.2a}) below):
In other words, 
the $Q^2$ evolution of the Mellin moment with certain value
$n$ does not depend on the moment with another value $n^\prime$.

The Mellin moments $M_k(n,Q^2)$ of the SF $F_k(x,Q^2)$
\begin{equation}
M_k(n,Q^2) = \int_{0}^{1}dx\,x^{n-2} F_k(x,Q^2)
\label{S1.3}
\end{equation}
can be represented
as the sum
\begin{equation}
M_k(n,Q^2) = \sum_{a=q,\bar{q},g} \, C^a_k(n,Q^2/\mu^2) \,
%\underbrace{C^a_k(n,Q^2/\mu^2)}_{\mbox{\LARGE\sf Coeff. function}}
A_a(n,\mu^2),
\label{S1.4}
\end{equation}
where $C^a_k(n,Q^2/\mu^2)$ are the coefficient functions and
$A_a(n,\mu^2)=<N| \mathcal{O}^a_{\mu_1, ..., \mu_n}|N>$ are the matrix
elements of the Wilson operators $\mathcal{O}^a_{\mu_1, ..., \mu_n}$,
which in turn are process independent.

Phenomenologically, the matrix elements $A_a(n,\mu^2)$ are equal to
the Mellin moments of the PDFs $f_a(x,\mu^2)$,
where $f_a(x,\mu^2)$ 
are the distributions\footnote{
All parton densities are multiplied by $x$, i.e. 
in the LO the
structure functions are some combinations of the parton densities.}
of quarks ($a=q_i$), antiquarks ($a=\bar{q}_i$) with
$i=1 ... 6$ and gluons ($a=g$), i.e.
\begin{equation}
A_a(n,\mu^2) \equiv f_a(n,\mu^2) = \int_{0}^{1}dx\,\,x^{n-2} f_a(x,\mu^2).
\label{S1.5}
\end{equation}
\noindent
The coefficient functions $C^a_k(n,Q^2/\mu^2)$ are represented by 
\begin{equation}
C^a_k(n,Q^2/\mu^2) = \int_{0}^{1}dx\,\,x^{n-2} \tilde{C}^a_k(x,Q^2/\mu^2)
\label{S1.6}
\end{equation}
\noindent
and responsible for the relationship between SFs and PDFs. Indeed, in
% because in
the $x$-space the relation (\ref{S1.4}) is replaced by
\begin{equation}
F_k(x,Q^2) = \sum_{a=q,\bar{q},g} \,
\tilde{C}^a_k(x,Q^2/\mu^2) \otimes f_a(x,\mu^2),
\label{S1.7}
\end{equation}
\noindent
where $\otimes$ denotes the Mellin convolution
\begin{equation}
f_1(x) \otimes f_2(x) \equiv \int^1_x \, \frac{dy}{y} \, f_1(y)
f_2\left(\frac{x}{y}\right).
\label{S1.8}
\end{equation}
\noindent
Applying~(\ref{S1.4}) and (\ref{S1.7}),
%of DIS process
one can fit
the shapes of PDFs $f_a(x,\mu^2)$, which are process-independent and
use them later for other processes. 
Note that the factorization scale  $\mu^2$ is often taken as $\mu^2=Q^2$. Here we will follow this choice.

\subsection{Quark densities
%  distribution functions
} \indent

%{\bf 2.}
The distributions of the $u$ and $d$ quarks contain the valence 
and the sea parts:
\begin{equation}
f_{q_1} \equiv f_u = f_u^V + f_u^S ~,~~ f_{q_2} \equiv f_d = f_d^V + f_d^S.
\label{S1.10}
\end{equation}
The distributions of the other quark flavors and of
all the antiquarks contain the sea parts only:
\begin{equation}
f_{q_j} = f_{q_j}^S,~~(j=3 ... 6),~~~ f_{\bar{q}_i} = f_{\bar{q}_i}^S ~~(i=1 ... 6).
\label{S1.11}
\end{equation}
\noindent
It is useful to define the combinations\cite{Buras} of quark densities\footnote{Here
we consider all quark flavors. Really, heavy quarks factorize out when $\sqrt{Q^2}$ becomes
less then their masses, and we should exclude them from the $Q^2$-region.},
the valence part $f_{V}$, the sea one $f_{S}$ and the singlet one $f_{SI}$:
%Following to \cite{Buras}, it is possible to introduce the useful
%%following
%combinations of quark densities: the valence part $f_{V}$, the sea one
% $f_{S}$ and the singlet one $f_{SI}$
\begin{eqnarray}
f_{V} &=& f_u^V + f_d^V ~,~~
f_{S} = \sum_{i=1}^{6} \left(f_{q_i}^S + f_{\bar{q}_i}^S \right),
\nonumber \\
f_{SI} &=& \sum_{i=1}^{6} \left(f_{q_i} + f_{\bar{q}_i} \right) = f_{V} + f_{S}.
\label{S1.12}
\end{eqnarray}
\noindent
Because the PDFs, which contribute to the structure functions, 
are accompanied by some numerical factors, there are also nonsinglet parts
\begin{equation}
f_{\Delta_{ij}} = \left(f_{q_i} +f_{\bar{q}_i}\right)-
 \left(f_{q_j} +f_{\bar{q}_j}\right),
\label{S1.13}
\end{equation}
which contain difference of densities of quarks and antiquarks with
different values of charges.

As an example, we consider the electron-proton scattering, where the
corresponding SF has the form
\begin{equation}
F_2^{ep}(x,Q^2) = \sum_{i=1}^{6} e_i^2  \left(f_{q_i}(x,Q^2) +
f_{\bar{q}_i}(x,Q^2) \right).
\label{S1.14}
\end{equation}
In the four-quark case (when $b$ and $t$ quarks are separated out), we will have\cite{Buras}
\begin{equation}
F_2^{ep}(x,Q^2) = \frac{5}{18} \, f_{SI}(x,Q^2) + \frac{1}{6} f_{\Delta}(x,Q^2),
\label{S1.15}
\end{equation}
where
\be
%\begin{align}
f_{\Delta}
%&
= \sum_{q_i=u,c} \left(f_{q_i}(x,Q^2) +
f_{\bar{q}_i}(x,Q^2) \right)
%\nonumber \\&
- \sum_{q_i=d,s} \left(f_{q_i}(x,Q^2) +
f_{\bar{q}_i}(x,Q^2) \right).
\label{S1.16}
%\end{align}
\ee

\subsection{DGLAP equations
  %and its solution
} \indent

%{\bf 3.}~~
The PDFs obey the DGLAP equations\cite{DGLAP}:
\bea
&&\frac{d}{d\ln {Q^{2}}} \, f_{i}(x,Q^{2}) ~=
%\int_{x}^{1}\frac{dy}{y}
-\frac{1}{2} \, \sum_{b} \gamma_{NS}(x) \otimes f_{i}(x,Q^{2}),~~i=NS,V,, \nonumber \\
&&\frac{d}{d\ln {Q^{2}}} \, f_{a}(x,Q^{2}) ~=
%\int_{x}^{1}\frac{dy}{y}
-\frac{1}{2} \, \sum_{b} \gamma_{ab}(x) \otimes f_{b}(x,Q^{2}),~~a,b=SI,g,
\label{S2.1}
\eea
where $\gamma_{i}(x)$
%$a,b=NS,SI,g$
and $\gamma_{ab}(x)$ are the so-called
%corresponding
splitting functions.
Anomalous dimensions (ADs) $\gamma_{ab}(n)$ of the twist-two Wilson operators
$\mathcal{O}^a_{\mu_1, ..., \mu_n}$ in the brackets $b$
%(hereafter $a_s=\alpha_s/(4\pi)$)
are the Mellin transforms of the corresponding
splitting functions
\begin{equation}
\gamma _{ab}(n) = \int_{0}^{1}dx\,\,x^{n-2} \gamma_{ab}(x),~~  f_a(n,\mu^2) = \int_{0}^{1}dx\,\,x^{n-2} f_a(x,\mu^2).
%~=~ \sum_{m=0}^{\infty} \gamma^{(m)} _{ab}(j) a_s^m ,
\label{S2.2}
\end{equation}
\noindent
At the LO of perturbation theory, ADs
%the anomalous dimensions
$\gamma_{ab}(n)$ have the following form\cite{Buras}:
\bea
&&\gamma_{ab}(n)= a_s(Q^2) \, \gamma^{(0)}_{ab}(n),~~ \gamma^{(0)}_{NS}(n)=\gamma^{(0)}_{qq}(n),~~ a_s(Q^2)=\frac{\alpha_s(Q^2)}{4\pi}
= \frac{1}{\beta_0\ln(Q^2/\Lambda^2_{\rm LO})},~~\nonumber \\
%\label{LO} \\
&&\gamma^{(0)}_{NS}(n)=\gamma^{(0)}_{qq}(n)=8C_F \left(S_1(n)-\frac{3}{4} - \frac{1}{2n(n+1)}\right), \nonumber \\
%\label{NS}\\
&&\gamma^{(0)}_{qg}(n)=-4f \frac{n^2+n+2}{n(n+1)(n+2)},~~\gamma^{(0)}_{gq}(n)=-4C_F \frac{n^2+n+2}{n(n^2-1)}, \nonumber \\
%\label{qg}\\
&&\gamma^{(0)}_{gg}(n)=8C_A \left(S_1(n)-\frac{11}{12} - \frac{1}{n(n-1)}- \frac{1}{(n+1)(n+2)}\right)+ \frac{4f}{3},  \label{gg}
\eea
where $C_A=N$, $C_F=(N^2-1)/(2N)$ for SU(N) group and $f$ is the number of active (massless) quarks and
\be
S_1(n)=\sum_{m=1}^{\infty} \, \frac{1}{m} = \Psi(n+1) + \gamma_{E},
\label{S1}
\ee
with Euler $\Psi$-function and Euler constant $\gamma_{E}$.

In the Mellin moment space, the DGLAP equation becomes to be
the standard renormalization group equation. At LO we have
\bea
&& \frac{d}{d\ln {Q^{2}}} \, f_{i}(n,Q^{2}) ~= -\frac{a_s(Q^2)}{2} \,
%\int_{x}^{1}\frac{dy}{y}
\gamma^{(0)}_{NS}(n)\, f_{i}(n,Q^{2}),~~i=V,NS,
%\nonumber \\
\label{S2.NS}  \\
 && \frac{d}{d\ln {Q^{2}}} \, f_{a}(n,Q^{2}) ~= -\frac{a_s(Q^2)}{2}
%\int_{x}^{1}\frac{dy}{y}
\sum_{b=SI,G} \gamma^{(0)}_{ab}(n)\, f_{b}(n,Q^{2}),~~ a=SI,g.
\label{S2.2a}
\eea
%where
%\be
%a_s(Q^2) \equiv \frac{\alpha_s(Q^2)}{4\pi} = \frac{1}{\beta_0\ln(Q^2/\Lambda^2_{\rm LO})}\,.
%\label{as}
%\ee

\noindent
To solve~(\ref{S2.2a}), it is better to move to the $\pm$ components\cite{Buras,Kotikov:2015nsa}, that leads to the diagonal form:
\be
\frac{d}{d\ln {Q^{2}}} \, f_{\pm}(n,Q^{2}) ~= -\frac{a_s(Q^2)}{2}
%\int_{x}^{1}\frac{dy}{y}
\gamma^{(0)}_{\pm}(n)\, f_{\pm}(n,Q^{2}),
%~~ a=S,g,
\label{S2.pm}
\ee
where
\be
\gamma^{(0)}_{\pm}(n) =\frac{1}{2} \, \Bigl[\gamma^{(0)}_{qq}(n)+\gamma^{(0)}_{gg}(n) \pm \sqrt{(\gamma^{(0)}_{qq}(n)-\gamma^{(0)}_{gg}(n))^2
    +4\gamma^{(0)}_{qg}(n)\gamma^{(0)}_{gq}(n)}\Bigr].
  \label{ga.pm}
  \ee
\noindent
The solutions of~(\ref{S2.NS}) and (\ref{S2.pm}) have the following form:
\be
f_{a}(n,\mu^{2})=f_{a}(n,Q_0^{2}) \, e^{-d_a(n) s},~~ a=V,NS,\pm, \label{Solu}
\ee
where $Q_0^2$ is some initial scale and
\be
d_a(n)=\frac{\gamma^{(0)}_{a}(n)}{2\beta_0},~~s=\ln \frac{\ln (Q^2/\Lambda^2)}{\ln (Q_0^2/\Lambda^2)}\,.~
\label{d_a}
\ee
\noindent
The singlet quark and gluon densities can be expressed through their ``$\pm$'' components as
\be
f_{a}(n,Q^{2})= f_{a,+}(n,Q^{2})+f_{a,-}(n,Q^{2}),~~
%\nonumber \\&&
f_{a,\pm}(n,Q^{2})=f_{a,\pm}(n,Q_0^{2}) \, e^{-d_{\pm}(n) s}, \label{Solu_a}
\ee
where
\bea
&&f_{q,+}(n,Q_0^{2})=f_{q}(n,Q_0^{2})-f_{q,-}(n,Q_0^{2}),~f_{q,-}(n,Q_0^{2})= f_{q}(n,Q_0^{2})\,\alpha_n + f_{q}(n,Q_0^{2})\beta_n,  \nonumber \\
&&f_{g,-}(n,Q_0^{2})=f_{g}(n,Q_0^{2})-f_{g,+}(n,Q_0^{2}),~f_{q,+}(n,Q_0^{2})= f_{g}(n,Q_0^{2})\,\alpha_n - f_{q}(n,Q_0^{2})\ep_n, \label{f0_a_pm}
\eea
and
\be
%\bea&&
\alpha_n=\frac{\gamma^{(0)}_{qq}(n)-\gamma^{(0)}_{+}(n)}{\gamma^{(0)}_{-}(n)-\gamma^{(0)}_{+}(n)},~~
\beta_n=\frac{\gamma^{(0)}_{qg}(n)}{\gamma^{(0)}_{-}(n)-\gamma^{(0)}_{+}(n)},~~
%\nonumber \\&&
\ep_n=\frac{\gamma^{(0)}_{gq}(n)}{\gamma^{(0)}_{-}(n)-\gamma^{(0)}_{+}(n)}.\label{alpha_n}
%\eea
\ee

\subsection{Special cases} \indent

Special case of parton evolution is the case $n=1$ for the valence part,
%and nonsinglet one,
which corresponds to number $N_V$ of structural quarks in considered hadron.
For example, for proton $N_V=3$. So, we have
\be
\int_0^1 dx \, \frac{1}{x} f_{V}(x,\mu^{2})=N_V\,,
%f_{a}(n=1,\mu^{2})=f_{a}(n=1,\mu_0^{2}),~~ (a=V,NS)\,,
\label{Struc}
\ee
which is the so-called Gross-Llewellyn-Smith sum rule \cite{GrLl}.

\noindent
Indeed, for this case, $\gamma^{(0)}_{NS}(n=1)=0$ and
\be
f_{a}(n=1,Q^{2})=f_{a}(n=1,Q_0^{2}),~~ a=V,NS.
\label{Solu_n1}
\ee
\noindent
For the NS part, the corresponding sum rule, so-called Gottfried sum rule \cite{Gottfried:1967kk}, is called as
\be
\int_0^1 dx \, \frac{1}{x} f_{NS}(x,Q^{2})=N_{NS}(Q^2)=3I_G(Q^2)\,,
%f_{a}(n=1,\mu^{2})=f_{a}(n=1,\mu_0^{2}),~~ (a=V,NS)\,,
\label{Gott}
\ee
with \cite{NewMuon:1993oys}
\be
I_G(Q_c^2)=0.705 \pm 0.078,~~~ Q_c^2=4\, \mbox{GeV}^2\,.
\label{GottQc}
\ee
\noindent
We note that the result (\ref{Gott}) is correct in the case of flavor-symmetric sea. Moreover,
  %{\bf Really,
$I_G(Q^2)$ has only very weak $Q^2$-dependence (see \cite{Kotikov:2005gr}), which comes
beyond LO from the so-called analytic continuation
  \cite{Kotikov:2005gr,KaKo} of the corresponding Wilson coefficients. So,
  %and, thus,
  the values of the Gottfried sum rule \cite{Gottfried:1967kk} can be taken below as
\be
I_G(Q^2) \approx I_G(Q_c^2)=0.705\,.
\label{GottMu}
\ee

\noindent
For the sea quark and gluon densities the special case is the $n=2$, that 
corresponds to the conservation of total momentum carried by quarks and gluons, i.e.
%to saving the total momentum carry out by quarks and gluons
%the total momentum of quarks and gluons, i.e.
\be
\int_0^1 dx \, \Bigl(f_{SI}(x,Q^{2})+f_{g}(x,Q^{2})\bigr)= \int_0^1 dx \, \Bigl(f_{SI}(x,Q_0^{2})+f_{g}(x,Q_0^{2})\bigr)=1,
%f_{a}(n=1,\mu^{2})=f_{a}(n=1,\mu_0^{2}),~~ (a=V,NS)\,,
\label{TotMom}
\ee
i.e.
\be
f_{SI}(n=2,Q^{2})+f_{g}(n=2,Q^{2}) = f_{SI}(n=2,Q_0^{2})+f_{g}(n=2,Q_0^{2})\bigr)=1.
\label{TotMom.1}
\ee
\noindent
Consider the case $n=2$ more accurately. We have
\be
\gamma^{(0)}_{qq}(n=2)=-\gamma^{(0)}_{gq}(n=2)=\frac{16C_F}{3},~~\gamma^{(0)}_{gg}(n=2)=-\gamma^{(0)}_{qg}(n=2)= \frac{4f}{3},
\label{ab.n2}
\ee
and, thus,
\be
\gamma^{(0)}_{-}(n=2)=0,~~\gamma^{(0)}_{+}(n=2)=\frac{4}{3}\Bigl(4C_F+f\Bigr),~~
%\label{pm.n2}\\&&
\alpha_{n=2}=\beta_{n=2}=1-\ep_{n=2}= \frac{f}{4C_F+f}. \label{alpha.n2}
\ee
\noindent
Using these values, we obtain
\bea
f_{SI,-}(2,Q^{2})&=&\frac{f}{4C_F+f}\, \Bigl(f_{SI}(2,Q_0^{2})+f_{g}(2,Q_0^{2})\Bigr)\, e^{-d_{-}(n=2)s} =\frac{f}{4C_F+f},~~\nonumber \\
f_{g,-}(2,Q^{2})&=&\frac{4C_F}{4C_F+f}\, \Bigl(f_{SI}(2,Q_0^{2})+f_{g}(2,Q_0^{2})\Bigr)\, e^{-d_{-}(n=2)s} =\frac{4C_F}{4C_F+f}, 
\,\label{fa.-}
\eea
because $f_{SI}(2,\mu_0^{2})+f_{g}(2,\mu_0^{2})=1$ and $d_{-}(n=2)=\gamma_{-}(n=2)/(2\beta_0)=0$.
Thus, the ``$-$''-components are $Q^2$-independent. Moreover,
\be
f_{SI,-}(2,Q^{2})+f_{g,-}(2,Q^{2})=f_{SI}(2,Q_0^{2})+f_{g}(2,Q_0^{2})=1, \label{Efa.-}
\ee
i.e. the sum of the ``$-$'' components of the singlet and gluon densities is responsible for the momentum conservation.
%is $Q^2$-independent,  
For the ``$+$'' components we have
\bea
&&f_{SI,+}(n,Q^{2})=\frac{1}{4C_F+f}\, \Bigl(4C_F \,f_{SI}(2,Q_0^{2})- f\,f_{g}(2,Q_0^{2})\Bigr)\, e^{-d_{+}(n=2)s}, \nonumber \\
&&f_{g,+}(n,Q^{2})= \frac{1}{4C_F+f}\, \Bigl(f\,f_{g}(2,Q_0^{2})-4C_F \,f_{SI}(2,Q_0^{2})\Bigr)\, e^{-d_{+}(n=2)s} \label{alpha.n2}
\eea
and, thus,
\be
f_{SI,+}(2,Q^{2})+f_{g,+}(2,Q^{2})=0, \label{Efa.+}
\ee
i.e. the  sum of the ``$+$'' components of the singlet and gluon densities is exactly zero.

%So, finally we have two relations, i.e. (\ref{Efa.-}) and (\ref{alpha.n2}), between ``$\pm$''-components. 

\section{Low $x$ asymptotics} \indent

According to~(\ref{S1.12}), singlet quark density $f_{SI}(x,Q^2)$ contains the
valence part $f_{V}(x,Q^2)$ and sea part $f_{S}(x,Q^2)$.

\subsection{Nonsinglet and valence parts} \indent

%{\bf 2.} ~
At small-$x$ values the NS and valence parts
%Here and below we will consider only the valent and nonsinglet parts
%%$(i=V,NS)$ of parton densities, which
have the following asymptotics\cite{Gross,LoYn}:
\begin{equation}
f_i(x) ~\to~  A_i(s) \, x^{\lambda_i}, ~~~i=V,NS,
\label{S3.1}
\end{equation}
where
%\begin{gather}
\be
%s=\ln \frac{\ln (Q^2/\Lambda^2)}{\ln (Q_0^2/\Lambda^2)},~
A_i(s) = A_i(0) e^{-d_{V}(1-\lambda_i) s},~
%\,,~~
A_i(0)\equiv  A_i,~
%\label{S3.2} \\
d_{V}(n) = \frac{\gamma^{(0)}_{NS}(n)}{2\beta_0} \,,
%%\frac{32}{3}
%\frac{16}{3\beta_0} \left[
%\Psi(n+1)+ \gamma_E -\frac{3}{4}- \frac{1}{2n(n+1)} \right].
\label{S3.2} 
\ee
%\nonumber
%\end{gather}
%where
$\lambda_i$ and $A_i(0)$ are free parameters
and $\Psi(n+1)$ is Euler $\Psi$-function.
>From the Regge calculus,
the constant $\lambda_i \sim 0.3 \div 0.5$.
Moreover,
%$\lambda_i$
the $Q^2$ evolution of this parton density shows that
$\lambda_i$ should be $Q^2$ independent \cite{LoYn}.

\subsection{Singlet part} \indent

It was pointed out\cite{BF1} that the HERA small-$x$ data can be
well interpreted in
terms of the so-called doubled asymptotic scaling (DAS) phenomenon
related to the asymptotic
behavior of the DGLAP evolution discovered many years ago\cite{Rujula}.
The study\cite{BF1} was extended\cite{Munich,Q2evo}
%\cite{Munich,Q2evo,HT}
to include the finite parts of anomalous dimensions 
of Wilson operators and Wilson coefficients\footnote{
In the standard DAS approximation\cite{Rujula} only the AD singular
parts
%of the anomalous dimensions
were used.}.
This led to predictions \cite{Q2evo} of the small-$x$ asymptotic
%PD
form of PDFs
in the framework of the DGLAP dynamics,
%equation
which were obtained
starting at some $Q^2_0$ with
the flat function
 \begin{equation}
f_a (x, Q^2_0) = A_a,
\label{1}
 \end{equation}
 where
 %$f_a$ are PDFs
%%the parton distributions
 %multiplied by $x$, $a = q$ or $g$ and
 $A_a$ are free parameters which have 
to be determined from the data.
%From now on, we
We refer to the approach of\cite{Munich,Q2evo}
%\cite{Munich,Q2evo,HT}
as {\it generalized} DAS approximation. In this approach
%generalized DAS
the flat initial conditions~(\ref{1}) determine the
basic role of the AD singular parts
%of anomalous dimensions,
as in the standard DAS case,
whereas the contributions coming from AD finite parts and Wilson
coefficients can be considered as corrections which are, however, important for
achieving
better agreement with experimental data.

Hereafter we consider for simplicity only  the LO
%leading order (LO)
approximation\footnote{Both the LO and NLO results and their applications can be found in\cite{Q2evo} and \cite{Kotikov:2016oqm}, respectively.}. 
The small-$x$ asymptotic expressions for sea quark and gluon densities $f_a(x, \mu^2)$
%(hereafter $\tilde{f}_q^S \equiv f_q$)
can be written as follows:
%--\cite{Cvetic1}):
\begin{eqnarray}
f_a(x,Q^2) &=&
%~=~
f_a^{+}(x,Q^2) + f_a^{-}(x,Q^2), \nonumber \\
f^{+}_g(x,Q^2) &=&
%\biggl(A_g + C \,
%%\frac{4}{9}
%A_q \biggl)
A_g^+	\overline{I}_0(\sigma) \; e^{-\overline d_{+} s} + O(\rho),~~ A_g^+= A_g + C \,A_q,~~
C=\frac{C_F}{C_A}=\frac{4}{9}, \nonumber \\
f^{+}_q(x,Q^2) &=& A_q^+
%%&=& \frac{f}{9}
%\frac{\varphi}{3} \,\biggl(A_g + C \,
%%\frac{4}{9}
%A_q \biggl)
%\rho
\tilde{I}_1(\sigma)  \; e^{-\overline d_{+} s}
+ O(\rho),~~A_q^+= \frac{\varphi}{3} \, A_g^+,~~
\varphi=\frac{f}{C_A}=\frac{f}{3},
%	\label{8.01} \\
\nonumber \\
f^{-}_g(x,Q^2) &=& A_g^{-} \,
%- C \,
%%        \frac{4}{9}
%A_q
e^{- d_{-} s} \, + \, O(x),~~ A_g^- =-C \,A_q,
        \nonumber \\
%	\label{8.00} \\
	f^{-}_q(x,Q^2) &=&  A_q e^{-d_{-} s} \, + \, O(x),
	\label{8.02}
\end{eqnarray}
\noindent
where $C_A=N_c$, $C_F=(N_c^2-1)/(2N_c)$ for the color $SU(N_c)$ group,
$\overline{I}_{\nu}(\sigma)$ and
%($\nu=0,1$)
$\tilde{I}_{\nu}(\sigma)$ ($\nu=0,1$)
are the combinations of the modified Bessel functions (at $s\geq 0$, i.e. $\mu^2 \geq Q^2_0$) and usual
Bessel functions (at $s< 0$, i.e. $\mu^2 < Q^2_0$):
\bea
%\begin{equation}
%\rho^{\nu}
&&\tilde{I}_{\nu}(\sigma) =
\left\{
\begin{array}{ll}
\rho^{\nu} I_{\nu}(\sigma) , & \mbox{ if } s \geq 0; \\
(-\tilde{\rho})^{\nu} J_{\nu}(\tilde{\sigma}) , & \mbox{ if } s < 0.
\end{array}
\right. \, ,~~
%\rho^{-\nu}
\overline{I}_{\nu}(\sigma) =
\left\{
\begin{array}{ll}
\rho^{-\nu} I_{\nu}(\sigma) , & \mbox{ if } s \geq 0; \\
\tilde{\rho}^{-\nu} J_{\nu}(\tilde{\sigma}) , & \mbox{ if } s < 0.
\end{array}
\right.
%\label{4}
\nonumber \\
%\end{equation}
&&I_{\nu}(\sigma)=\sum_{m=0}^{\infty} \, \frac{1}{k!(k+\nu)!}\, \sigma^{2k+\nu},~~J_{\nu}(\sigma)=\sum_{m=0}^{\infty} \, \frac{(-1)^k}{k!(k+\nu)!}\, \sigma^{2k+\nu},
\label{4}
\eea
where $\overline{I}_{0}(\sigma) = \tilde{I}_{0}(\sigma)$ and
\be
%\bea
%\begin{equation}&&
%&&
%s=\ln \frac{a_s(Q^2_0)}{a_s(\mu^2)},~~
%a_s(\mu^2) \equiv \frac{\alpha_s(\mu^2)}{4\pi} = \frac{1}{\beta_0\ln(\mu^2/\Lambda^2_{\rm LO})},~~
\sigma = 2\sqrt{\left|\hat{d}_+\right| s
  \ln \left( \frac{1}{x} \right)},~~
%\nonumber \\&&
\rho=\frac{\sigma}{2\ln(1/x)},~~
\tilde{\sigma} = 2\sqrt{-\left|\hat{d}_+\right| s
  \ln \left( \frac{1}{x} \right)},~~ \tilde{\rho}=\frac{\tilde{\sigma}}{2\ln(1/x)},
\label{intro:1a}
\ee
and
\begin{equation}
\hat{d}_+ = - \frac{4C_A}{\beta_0} = - \frac{12}{\beta_0},~~~
\overline d_{+} = 1 + \frac{4f(1-C)}{3\beta_0} =
1 + \frac{20f}{27\beta_0},~~~
d_{-} = \frac{4Cf}{3\beta_0}= \frac{16f}{27\beta_0},
\label{intro:1b}
\end{equation}
are the singular and regular parts of the anomalous dimensions
and $\beta_0 = 11 -(2/3) f$ is the first coefficient of the QCD
$\beta$-function in the $\MSbar$-scheme.
The results for the parameters $A_a$ and $Q_0^2$ can be found in \cite{Cvetic1};
%which have been
they were obtained
%\footnote{In the future, by using  (\ref{8a}) and (\ref{8.02}) and
%  % studies using the results (\ref{8a}) and (\ref{8.02}) and the ones
%  results of \cite{Illarionov:2008be}
%we plan to perform the combined fits to the H1 and ZEUS experimental data\cite{37} and\cite{Abramowicz:1900rp} for
%the DIS structure function $F_2(x,Q^2)$ and its charm part $F_2^c(x,Q^2)$, respectively.}
for $\alpha_s(M_Z)=0.1168$.

It is convenient to show the following expressions:
\begin{equation}
\beta_0 \, \hat{d}_+ = - 4C_A,~~~
\beta_0 \, \overline{d}_{+} = \frac{C_A}{3}\Bigl(11+2\varphi (1-2C)\Bigr),~~~
\beta_0 \, d_{-} = \frac{4Cf}{3}= \frac{4C_F \varphi}{3}.
\label{intro:1ba}
\end{equation}

%%%%%%%%%%%%%%%%%%%%%

\section{Large $x$ asymptotics} \indent

%According to~(\ref{S1.12}),
%singlet quark density $f_{SI}(x,Q^2)$ contains the
%valence part $f_{V}(x,Q^2)$ and sea part $f_{S}(x,Q^2)$.

%\subsection{Nonsinglet and Valence parts}

%{\bf 1.} ~
The large $x$ asymptotics of the valence, nonsinglet and sea quark densities have
the following form (see\cite{Gross,LoYn} and Appendix A):
\bea
\z  f_i(x,Q^2) \approx \frac{B_i(s)}{\Gamma(1+\nu_i(s))} (1-x)^{\nu_i(s)},~~i=NS,V, \nonumber \\
\z f_a(x,Q^2) = \sum_{\pm}\, f_{a,\pm}(x,Q^2),~~ a=q,g,~~ \nonumber \\
%\label{sg_l} \\
\z f_{q,-}(x,Q^2)  \approx \frac{B_{-}(s)\, }{\Gamma(1+\nu_{-}(s))}\,  (1-x)^{\nu_{-}(s)},\nonumber \\
%\label{q-l}\\
\z f_{g,-}(x,Q^2)  \approx \frac{K_{-}}{\Gamma(2+\nu_{-}(s))}  \,
  \frac{B_-(s) }{\left[\ln(1/(1-x))+\hat{c} + \Psi(\nu_{-}+2)\right]} 
\,  (1-x)^{\nu_{-}(s)+1}, \nonumber \\
%\label{g-l}\\
\z f_{g,+}(x,Q^2) \approx \frac{B_{+}(s)\, }{\Gamma(1+\nu_{+}(s))}\, (1-x)^{\nu_{+}(s)}, \nonumber \\
%\label{g+l}\\
\z f_{q,+}(x,Q^2) \approx  - \frac{ K_{+}}{\Gamma(2+\nu_{+}(s))}  \, \frac{B_+(s) x}{\left[\ln(1/(1-x))+\hat{c} + \Psi(\nu_{+}+2)\right]}  
  (1-x)^{\nu_{+}(s)+1}\,,
%  % + B_g(s)x^{\mu_{+}(s)}
\label{S2.3}
\eea
%where $K_{\pm}$ and $\hat{c}$ are shown in (\ref{Ki}).
where
%$(j=NS,V,S)$
\begin{gather}
%s=\ln \left(\frac{\ln (Q_0^2/\Lambda^2)}{\ln (Q^2/\Lambda^2)}\right) \,,~
\nu_i(s) = \nu_i(0) + r_{i} s \,,~
r_{i} = \frac{4C_i}{\beta_0} \,,~
%\nonumber \\
%\hat{d}_{G}= \frac{12}{\beta_0}, ~~
%\nonumber \\
B_i(s) =
%&=&
B_i(0)\,
%\frac{
e^{-p_i s}
%}{\Gamma(1+\nu_i(s))}
\,,~
%\nonumber \\
p_i =
%&=&
r_{i} \Bigl(\gamma_E+ \hat{c}_i
%\frac{3}{4}
\Bigr),~~i=NS,V,\pm,
%~~  p_G= \hat{d}_{G} \left(\gamma_E-\frac{\beta_0}{12}\right) s.
\label{S2.4}
%\nonumber
\end{gather}
%are coefficients from large $x$ asymptotics
with $B_i(0)$ and $\nu_i(0)$ being free parameters, 
$\gamma_E$ is the Euler constant and
\bea
&&C_+=C_A,~~\hat{c}_+=-\frac{\beta_0}{4C_A}= - \frac{11-2\varphi}{12},~~C_j=C_F,~~\hat{c}_j=-\frac{3}{4},~~j=V,NS,\nonumber \\
&& K_+=\frac{f}{2(C_A-C_F)},~~K_-=\frac{C_F}{2(C_A-C_F)},~~\hat{c}=\gamma_{E}+ \frac{C_A\hat{c}_+-C_F\hat{c}_-}{C_A-C_F}
\label{Cj}
%\nonumber
\eea
\noindent
The constant $\nu_i(0)$ can be estimated
%taken
from the quark counting rules \cite{schot} as
%{\color{green} (V.A.~Matveev et al., 1973),
%(S.J.~Brodsky et al., 1973,1995)}:
\begin{equation}
\nu_j(0)
%\sim  \beta_{NS}(0)
\sim 3,~~ j=V,NS,-,~~
\nu_+(0)=\nu_j(0)+1 \,.
%~~~  %\nonumber \\ \beta_G(0) ~ \sim ~ \beta_{NS}(0)+1 \sim 4,  ~~~
%%\nonumber \\\beta_S(0) ~ \sim ~ \beta_{G}(0)+1 \sim 5  \, .
\label{S2.5}
%\nonumber
%\nonumber
\end{equation}

%relation 1 leads to the smallness of the term 2
The relation $\nu_+(s)=\nu_-(s)+1$ leads to the smallness of the term  $f_{q,+}(x,Q^2)$. So, at large $x$ we have
\be
 f_{q,+}(x,Q^2) \approx 0 ~~~\mbox{and, thus,}~~ f_{q}(x,Q^2) \approx f_{q,-}(x,Q^2).
\label{q+}
\ee
\noindent
Moreover, the expressions~(\ref{S2.3}) and (\ref{S2.4}) demonstrate the fall
%decreasing
of the parton densities at large $x$ when $Q^2$ increases.

\section{Parametrizations
%Valent part
} \indent

Here we present parametrizations of the nonsinglet and singlet quark and gluon densities constructed similar 
to ones obtained earlier\cite{Illarionov:2010gy} in the valence case.

\subsection{Nonsinglet and valence parts} \indent

The nonsinglet and valence quark part $f_i(x,Q^2)$, where $i=V,NS$, 
can be represented in the following form\footnote{Similar studies were carried out 
  %  have been also done
  in\cite{Kotikov:1988ha} (see also the review \cite{Kotikov2007}).}:
\begin{equation}
%f_V(x,Q^2) &=& \overline{f}_V(x,Q^2) \cdot \left(1+ \sum^N_{k=1}
%\alpha_{k,V} x^k \right)~~~(N=1,2,3), \nonumber \\
f_i(x,Q^2) = 
\biggl[A_i(s)x^{\lambda_i}(1 -x) + \frac{B_i(s)\, x}{\Gamma(1+\nu_i(s))} + D_i(s)x (1 -x)  \biggr] \, (1-x)^{\nu_i(s)} \,,
\label{S4.1_l}
%\nonumber
\end{equation}
which is constructed as a combination of the small $x$ and large $x$
asymptotics and an additional term proportional to $D_i(s)$, which is subasymptotics in both these regions.
The $Q^2$-dependence of the parameters in (\ref{S4.1_l}) is given by~(\ref{S2.3})
and (\ref{S3.2}).
The $Q^2$-dependence of magnitude $D_i(s)$
is determined by the corresponding sum rules (see below).

\subsection{Sea and gluon parts} \indent

The sea and gluon parts can be represented as combinations of the $\pm$ terms:
\bea
\z f_j(x,Q^2) = \sum_{\pm}\, f_{j,\pm}(x,Q^2),~~ j=q,g,~~ \nonumber \\
%\label{sg_l} \\
\z f_{q,-}(x,Q^2)=
\biggl[A_{q}e^{- d_{-} s} (1-x)^{m_{q,-}} +   \frac{B_{-}(s)\, x}{\Gamma(1+\nu_{-}(s))}+ D_{-}(s)x (1 -x)  \biggr]\,  (1-x)^{\nu_{-}(s)},\nonumber \\
%\label{q-l}\\
\z f_{g,-}(x,Q^2)=
\biggl[A_{g}^{-}e^{- d_{-} s} (1-x )^{m_{g,-}} \nonumber\\
&& + \frac{K_{-}}{\Gamma(2+\nu_{-}(s))}  \,
  \frac{B_-(s) x}{\left[\ln(1/(1-x))+\hat{c} + \Psi(\nu_{-}+2)\right]} \biggr]  
\,  (1-x)^{\nu_{-}(s)+1}, \nonumber \\
%\label{g-l}\\
\z f_{g,+}(x,Q^2)= 
\biggl[A_{g}^+ \overline{I}_0(\sigma)e^{-\overline d_{+} s} (1-x)^{m_{g,+}} + \frac{B_{+}(s)\, x}{\Gamma(1+\nu_{+}(s))} + D_{+}(s)x (1 -x)  \biggr]\, (1-x)^{\nu_{+}(s)}, \nonumber \\
%\label{g+l}\\
\z f_{q,+}(x,Q^2)= 
%  \biggl[
A_{q}^+ \tilde{I}_1(\sigma)e^{-\overline d_{+} s}
%(1-x)^{m_{q,+}}\nonumber \\
%   && - \frac{ K_{+}}{\Gamma(2+\nu_{+}(s))}  \, \frac{B_+(s) x}{\left[\ln(1/(1-x))+\hat{c} + \Psi(\nu_{+}+2)\right]} \biggr] 
  (1-x)^{\nu_{+}(s)+m_{q,+}+1}
%  % + B_g(s)x^{\mu_{+}(s)}
  ,\label{q+l}
\eea
where $K_{-}$ and $\hat{c}$ are shown in (\ref{Cj}).
%{\bf It is better
%We note that it is better to use $m_{q,-}=m_{g,+}=2$ and  $m_{q,+}=m_{g,-}=1$. 
%In the case, small-$x$ asymptotics is supressed at large $x$ to
%compare with subasymptotics $\sim D_{...}$. 
%Moreover, in the case small-$x$ asymptotics contain same powers of the factor $(1-x)$ for quarks and gluons.
%}\\ 
We note that one can
set $m_{q,-}=m_{g,+}=2$ and $m_{q,+}=m_{g,-}=1$. 
In this case, small-$x$ asymptotics is suppressed for large $x$ for comparison with the subasymptotic behavior of
%is supressed at large $x$ to
%compare with subasymptotics
$\sim D_{\pm}(x)$.
Moreover, the small-$x$ asymptotics will 
contain the same powers of the factor $(1-x)$ for quarks and gluons.
%,
%and
%%{\bf !!! As in the previous section, it
%it is possible to put $K_{+}=0$, since the part $\sim K_{+}$ is strongly suppressed.
%if we would like to recover (\ref{S2.3}) from (\ref{q+l}) at $s=0$.
%!!!}\\{\bf

We would like to note that the valence quarks contribute to the ``$-$''-component but not to ``$+$''-one. So,
\bea
&&f_{SI}(x,Q^2)= f_{SI,-}(x,Q^2)+f_{SI,+}(x,Q^2),\nonumber \\
&&f_{SI,-}(x,Q^2)= f_{q,-}(x,Q^2)+f_{V}(x,Q^2),~~f_{SI,+}(x,Q^2)= f_{q,+}(x,Q^2).
  \label{SI+-l}
\eea
\noindent
The parameters involved in (\ref{S4.1_l}) --- (\ref{SI+-l}) can be fitted, for example, from the
comparison with known parametrizations of NNPDF group\cite{NNPDF4}
%of dynamical PDFs\cite{NNPDF}
%%\cite{Jimenez-Delgado:2014twa}
and/or taking into account the sum rules shown in the next section.

\subsection{Properties of parameterizations
%  Sum rules
} \indent

The obtained above parametrizations of parton distributions in a proton
should obey sum rules given by (\ref{Struc}) and (\ref{Gott}).

\subsubsection{Gross-Llewellyn-Smith and Gottfried sum rules} \indent

The additional relations between the parameters in (\ref{S4.1_l})
stems
%for $i=V$
from the LO Gross-Llewellyn-Smith 
%following 
sum rule \cite{GrLl} and Gottfreed sum rule \cite{Gottfried:1967kk}:
%\footnote{Above LO, the Gross-Llewellyn-Smith sum rule \cite{GrLl} is
%defined as the integral of the structure function $F_3$ and contains
%the perturbative ($\sim \alpha_s$) and the power corrections in its r.h.s.
%(see, for example, \cite{KaSi} and references therein).}
\be
\int_0^1 \frac{dx}{x} f_i(x,Q^2) = N_i,~~i=V,NS,~ N_V=3,~N_{NS}=3I_G,
%\nonumber \\&&\int_0^1 \frac{dx}{x} f_{NS}(x,Q^2) = I_G(Q^2),
\label{S4.2l}
\ee
\noindent
where the value of $I_G$ can be found in~(\ref{GottMu}).

So, we have the following relations:
%(for $\alpha_{k,V}=0$)
\be
N_i = A_i(s)\frac{\Gamma(\lambda_i)\Gamma(2+\nu_i(s))}{\Gamma(\lambda_i+2+\nu_i(s))}
+ \frac{B_i(s)}{\Gamma(2+\nu_i(s))}+ \frac{D_i(s)}{2+\nu_i(s)},
%~~(i=V,NS)\,,
%\nonumber \\
% &&= A_i(s)\frac{\Gamma(\lambda_i)\Gamma(1+\nu_i(s))}{\Gamma(\lambda_i+1+\nu_i(s))} + \Bigl(B_i(s)-A_i(s)\Bigr)
% \frac{\Gamma(\mu_i(s))\Gamma(1+\nu_i(s))}{\Gamma(\mu_i(s)+1+\nu_i(s))} \,.
\label{SR_l}
\ee
i.e.
\be
D_i(s)= (2+\nu_i(s))\biggl[N_i -  A_i(s)\frac{\Gamma(\lambda_i)\Gamma(2+\nu_i(s))}{\Gamma(\lambda_i+2+\nu_i(s))} - \frac{B_i(s)}{\Gamma(2+\nu_i(s))}\biggr]\,.
\label{Di}
\ee
%where we used the notation $N_{NS}=I_G(Q^2)$.\\
\noindent
The valence and NS densities at low and large $x$ asymptotics are proportional each other.
  %in the NS and valence case.
  So, we can apply the following notations:
\be
\nu_V(s) \approx \nu_{NS}(s),~~\lambda_V \approx \lambda_{NS}
%\equiv \lambda
\,.
\label{VNS}
\ee

\subsubsection{Momentum conservation} \indent

The momentum conservation (\ref{TotMom}) leads to the following relations:
\be
1= G_V(s)+G^{-}_q(s)+G^{-}_g(s)\,~~
%\nonumber \\
0=G^{+}_q(s)+G^{+}_g(s),
\label{MC_pm_l}
\ee
where
\be
  \int_0^1 \,dx\, f_i(x,Q^2) = G_i(s),~~(i=V,NS),~~  \int_0^1 \,dx\, f_{a,\pm}(x,Q^2) = G^{\pm}_a(s),~~(a=q,g) .
\label{MC_int_l}
\ee
\noindent
So, we have
\bea
%\z G^{-}_s(s) = \frac{A_{q}}{1+\nu_{-}(s)+m_{q,-}}\, e^{-d_{-}s} +\frac{B_{-}(s)}{\Gamma(3+\nu_{-}(s))} + D_{-}(s)\frac{\Gamma(3/2)\Gamma(2+\nu_{-}(s))}{\Gamma(7/2+\nu_{-}(s))}
%, \nonumber \\
%%\label{G-s_l}\\
%\z G^{-}_g(s) = \frac{A^{(-)}_{g}}{2+\nu_{-}(s)+m_{g,-}}\, e^{-d_{-}s}
%%\nonumber \\&&
%+ K_{-}\, \frac{B_{-}(s)}{\Gamma(4+\nu_{-}(s))\left(\Psi(4+\nu_{-})+\hat{c}\right)},  \nonumber \\
%\label{G-g}
\z G_i(s) = A_i(s)\frac{\Gamma(\lambda_i+1)\Gamma(2+\nu_i(s))}{\Gamma(\lambda_i+3+\nu_i(s))}
% \nonumber \\&&
+ \frac{B_i(s)}{\Gamma(3+\nu_i(s))} +   \frac{D_i(s)}{(2+\nu_i(s))(3+\nu_i(s))}
%\nonumber \\
%&&= \frac{A_{SI,-}(s)}{1+\nu_{-}(s)}+ \Bigl(B_{SI}(s)-A_{SI,-}(s)\Bigr)\, \frac{\Gamma(\mu_{-}(s)+1)\Gamma(1+\nu_{-}(s))}{\Gamma(\mu_{-}(s)+2+\nu_{-}(s))}
 \,, \nonumber \\
%\label{V-l}\\
\z G^{+}_g(s) = A^{+}_{g}\Phi_{0}(m_{g,+}+\nu_{+}(s))\,e^{-\overline{d}_{+}s}+ \frac{B_+(s)}{\Gamma(3+\nu_{+}(s))}+ D_{+}(s)\frac{\Gamma(3/2)\Gamma(2+\nu_{+}(s))}{\Gamma(7/2+\nu_{+}(s))}
, \nonumber \\
\z G^{-}_q(s) = \frac{A_{q}}{1+\nu_{-}(s)+m_{q,-}}\, e^{-d_{-}s}
+\frac{B_{-}(s)}{\Gamma(3+\nu_{-}(s))}
+ D_{-}(s)\frac{\Gamma(3/2)\Gamma(2+\nu_{-}(s))}{\Gamma(7/2+\nu_{-}(s))}
\,, \nonumber \\
%\label{G-s_l.1}\\
%\z G^{-}_g(s) = \frac{A^{(-)}_{g}}{2+\nu_{-}(s)+m_{g,-}}\, e^{-d_{-}s}
%\,, \nonumber \\
%\label{G-s_l.1}\\
\z G^{-}_g(s) = \frac{A^{(-)}_{g}}{2+\nu_{-}(s)+m_{g,-}}\, e^{-d_{-}s}
+  \frac{K_{-}\,B_{-}(s)}{\Gamma(4+\nu_{-}(s))\left(\Psi(4+\nu_{-})+\hat{c}\right)}\,,~~\nonumber \\
\z G^{+}_q(s) = A^{+}_{q}\Phi_{1}(1+m_{q,+}+\nu_{+}(s)) \,e^{-\overline{d}_{+}s}
%\nonumber \\ &&
%- K_{+}\, \frac{B_{+}(s)}{\Gamma(4+\nu_{+}(s))\left(\Psi(4+\nu_{+})+\hat{c}\right)}
\,.
\label{G+s}
\eea
%where $\Phi_{j}(0,\nu_{+}(s))$ $(j=0,1)$ are given in Eq. (\ref{Phi0}) and  (\ref{Phi1}).
where 
\bea
%&&A^{-}_{S}=A_q+A_g^{-}=A_q (1-C), \nonumber \\
&&\Phi_{0}(\nu(s))=\int^1_0 \,dx\, I_0(\sigma) (1-x)^{\nu(s)} =\sum_{l=0}^{\infty} \, C_l^{\nu} \, \frac{(-1)^l}{l+1}
e^{(ds)/(l+1)}, \nonumber \\
%\label{Phi0}\\
&&\Phi_{1}(\nu(s))=\int^1_0 \,dx\, \rho\,I_1(\sigma) (1-x)^{\nu(s)} =\sum_{l=0}^{\infty} \, C_l^{\nu} \, (-1)^l e^{(ds)/(l+1)}, \label{Phi1}
\eea
with
\be
C_l^{\nu}=\frac{\Gamma(\nu+1)}{l!\Gamma(\nu+1-l)},~~ d=|\hat{d}_+|,
\label{hd}
\ee
and $\hat{d}_+<0$ is defined above in (\ref{intro:1ba}).
For $\nu=1, 2$ and $3$, we have:
\bea
&&\Phi_{j}(1)=
%\int^1_0 \, \overline{I}_0(\sigma)x^{\mu} (1-x)^{\nu} =
e^{ds}-\frac{1}{2-j} e^{ds/2}\,, ~~ j=0,1, \nonumber \\
&&\Phi_{j}(2)=
%\int^1_0 \, \overline{I}_0(\sigma)x^{\mu} (1-x)^{\nu} =
e^{ds}-(1+j) e^{ds/2}+\frac{1}{3-2j} e^{ds/3}, \nonumber \\
&&\Phi_{j}(3)=
%\int^1_0 \, \overline{I}_0(\sigma)x^{\mu} (1-x)^{\nu} =
e^{ds}- \frac{3}{2-j} e^{ds/2} +(1+2j) e^{ds/3}-\frac{1}{4-3j} e^{ds/4}
\,. \label{Phi13}
\eea
\noindent

We would like to note that comparing (\ref{Di}) and (\ref{G+s}) at $i=V,NS$ we have
\be
G_i(s) =  \frac{1}{3+\nu_i(s)}\, \biggl[N_V -  (1-\lambda_i)\,A_i(s)\frac{\Gamma(\lambda_i)\Gamma(3+\nu_i(s))}{\Gamma(\lambda_i+3+\nu_i(s))}
  + \frac{B_i(s)}{\Gamma(3+\nu_i(s))}\biggr]\,.
\label{Gv_dop}
\ee
Moreover,
\bea
\z D_{-}(s)=\frac{\Gamma(7/2+\nu_{-}(s))}{\Gamma(3/2)\Gamma(2+\nu_{-}(s))} \Bigl[1- G_V(s)-G^{-}_g(s)-\overline{G}^{-}_q(s)\Bigr]\,, \nonumber \\
%\label{D-s_l}\\
\z D_{+}(s)=-\frac{\Gamma(7/2+\nu_{+}(s))}{\Gamma(3/2)\Gamma(2+\nu_{+}(s))} \Bigl[G^{+}_q(s)+\overline{G}^{+}_g(s)\Bigr]\,,
\label{D+g_l}
\eea
where
%\be
\bea \z
\overline{G}^{-}_q(s) = \frac{A_{q}}{1+\nu_{-}(s)+m_{q,-}}\, e^{-d_{-}s}
+\frac{B_{-}(s)}{\Gamma(3+\nu_{-}(s))}\,, \nonumber \\
%\label{oG-s_l}\\
\z
\overline{G}^{+}_g(s) = A^{+}_{g}\Phi_{0}(m_{g,+}+\nu_{+}(s))\,e^{-\overline{d}_{+}s}+ \frac{B_+(s)}{\Gamma(3+\nu_{+}(s))}\,. \label{oG+g_l}\
\eea
\noindent
If the argument $\nu$ of $\Phi_{k}(\nu)$ is large, which is the case (see Section 5.4. below), then we have the approximation
%that is the case (see section 5.4. below) then we have an approximation
\bea
\z \Phi_{0}(\nu) \approx \frac{1}{1+\nu}\,  I_0(\sigma_{\nu}),~~ \sigma_{\nu}=\sigma ~~\mbox{with}~~\ln(1/x) \to \Psi(2+\nu)+\gamma_{\rm E} \approx \ln(1+\nu)+\gamma_{\rm E}\,, \nonumber \\
\z \Phi_{1}(\nu) \approx \frac{\rho_{\nu}}{1+\nu}\, I_1(\sigma_{\nu}),~ \rho_{\nu}=\rho ~\mbox{with}~\ln(1/x) \to \Psi(2+\nu)+\gamma_{\rm E} \approx \ln(1+\nu)+\gamma_{\rm E}\,,
%\nonumber \\\nonumber \\
\label{Phik}\
\eea
where $\sigma$ and $\rho$ are given in section 3.2 and $\gamma_{\rm E}$ is the Euler constant. 
The evaluating the results (\ref{Phik}) can be found in Appendix B.
Then, at $s=0$ we will have relations between $A^{\pm}_a$, $B_{\pm}(0)$ and $D_{\pm}(0)$:
  %. For example, in the case
  %$\lambda=1/2$,  $\mu^{(0)}_{V} = 1$ and $\mu^{(0)}_{\pm} = 1$
     %     $\nu_{-}(0)=3$ and $\nu_+(0)=4$
   %  we have (\ref{MC_pm.0}) and}
   \bea
%  \z G^{-}_s(s=0) = \frac{A_{q}}{1+\nu_{-}(0)+m_{q,-}} + \frac{B_{-}(0)}{\Gamma(3+\nu_{-}(0))} 
%+ D_{-}(0)\frac{\Gamma(3/2)\Gamma(2+\nu_{-}(0))}{\Gamma(7/2+\nu_{-}(0))}
%\,, \nonumber \\
%\z G^{-}_g(s=0) = \frac{A^{-}_{g}}{2+\nu_{-}(0)+m_{g,-}}+  \frac{K_{-}}{\Gamma(4+\nu_{-}(0))}\, \frac{B_{-}(0)}{\left(\Psi(4+\nu_{-}(0))+\hat{c}\right)}\,, 
%\nonumber \\
\z G_i(s=0) = \frac{1}{3+\nu_V(0)}\, \biggl[N_i -  (1-\lambda_V)\,A_i(0)\frac{\Gamma(\lambda_V)\Gamma(3+\nu_V(0))}{\Gamma(\lambda_V+3+\nu_V(0))}
  + \frac{B_i(0)}{\Gamma(3+\nu_V(0))}\biggr]\,,
\nonumber \\
%\label{MC-.s0_l}\\
\z G^{+}_g(s=0)= \frac{A^{+}_{g}}{1+\nu_{+}(0)+m_{g,+}} + \frac{B_{+}(0)}{\Gamma(3+\nu_{+}(0))} + D_{+}(0)\frac{\Gamma(3/2)\Gamma(2+\nu_{+}(0))}{\Gamma(7/2+\nu_{+}(0))}
\,, \nonumber \\
%\z G^{+}_s(s=0)=-  \frac{K_{+}}{\Gamma(4+\nu_{+}(0))} \, \frac{B_{+}(0)}{\left(\Psi(4+\nu_{+}(0))+\hat{c}\right)},
%\label{MC+.s0-l}
%\eea
%
%{\bf When $K_+=0$ and $B_{-}(s)=0$, we have for $G^{-}_s(s=0)$, $G^{-}_g(s=0)$
%  %in (\ref{MC-.s0_l})
%  and $G^{+}_s(s=0)$ in (\ref{MC+.s0-l})}
%  \bea
  \z G^{-}_q(s=0) = \frac{A_{q}}{1+\nu_{-}(0)+m_{q,-}}
  + \frac{B_{-}(0)}{\Gamma(3+\nu_{-}(0))} 
+ D_{-}(0)\frac{\Gamma(3/2)\Gamma(2+\nu_{-}(0))}{\Gamma(7/2+\nu_{-}(0))}
\,, \nonumber \\
\z G^{-}_g(s=0) = \frac{A^{-}_{g}}{2+\nu_{-}(0)+m_{g,-}}
+  \frac{K_{-}}{\Gamma(4+\nu_{-}(0))}\, \frac{B_{-}(0)}{\left(\Psi(4+\nu_{-}(0))+\hat{c}\right)},~  G^{+}_q(s=0)=0.~~
\label{MC+.s0-l.1}
\eea
\noindent
%{\bf When $\nu_{-}(0)=3$ and $\nu_+(0)=4$ we have}
So,
%  %from (\ref{MC-.s0_l}) and
%  (\ref{MC+.s0-l}) we have relations between $A_q$, $B_{-}(0)$ and $D_{-}(0)$ and also $A_g$, $B_{+}(0)$ and $D_{+}(0)$,
  %  respectively.}{\bf The
  the final results for $A_q$, $A_g$ and $B_{+}(0)$ can be obtained from experimental data for sea quark
  and gluon densities at $Q^2=Q_0^2$ (i.e. for $s=0$):
\bea 
\z f_j(x,Q_0^2) = \sum_{\pm}\, f_{j,\pm}(x,Q_0^2),~~ j=q,g,  \nonumber \\
%\label{sg} \\
%\z f_{q,-}(x,Q_0^2)=
%\biggl[A_{q}(1-x )^{m_{q,-}} +  \frac{B_-(0)\, x}{\Gamma(1+\nu_{-}(0))}  + D_{-}(0)x (1 -x)\biggr]\,  (1-x)^{\nu_{-}(0)} \,, \nonumber \\
%%\label{q-}\\
%\z f_{g,-}(x,Q_0^2)= \biggl[A_{g}^{-}(1-x )^{m_{g,-}}
%  %\nonumber \\&&
%  + \frac{K_{-}}{\Gamma(2+\nu_{-}(0))} \,
%  \frac{B_-(0) x}{\left[\ln(1/(1-x))+\hat{c} + \Psi(\nu_{-}(0)+2)\right]} \biggr]
%(1-x)^{\nu_{-}(0)+1} \,, \nonumber \\
%\label{g-}\\
\z f_{g,+}(x,Q_0^2)= 
\biggl[A_{g}^+ (1-x )^{m_{g,+}} + \frac{B_+(0)\, x}{\Gamma(1+\nu_{+}(0))} + D_{+}(0)x (1 -x)\biggr]\, (1-x)^{\nu_{+}(0)} \,,  \nonumber \\
%\label{g+}\\
%\z f_{q,+}(x,Q_0^2)= - \frac{K_{+}}{\Gamma(2+\nu_{+}(0))} \,
%  \frac{B_+(0) x}{\left[\ln(1/(1-x))+\hat{c} + \Psi(\nu_{+}(0)+2)\right]} % \biggr]
%(1-x)^{\nu_{+}(0)+1} 
%%A_{q}^+ \tilde{I}_1(\sigma)(1-x^{\mu_{+}(s)} )(1-x)^{\nu_{+}(s)}
%%  % + B_g(s)x^{\mu_{+}(s)}
\z f_{q,-}(x,Q_0^2)=
\biggl[A_{q}(1-x )^{m_{q,-}}
  +  \frac{B_-(0)\, x}{\Gamma(1+\nu_{-}(0))}
  + D_{-}(0)x (1 -x)\biggr]\,  (1-x)^{\nu_{-}(0)} \,,  \nonumber \\
%\label{q-.1}\\
\z f_{g,-}(x,Q_0^2)=
\biggl[
A_{g}^{-}
(1-x )^{m_{g,-}}
%\nonumber \\&&
+ \frac{K_{-}}{\Gamma(2+\nu_{-}(0))} \,
  \frac{B_-(0) x}{\left[\ln(1/(1-x))+\hat{c} + \Psi(\nu_{-}(0)+2)\right]} \biggr]
\nonumber \\&& \cdot (1-x)^{\nu_{-}(0)+1} \,,~~~
%\label{g-.1}\\&&
f_{q,+}(x,Q_0^2)=0
   \,,\label{q+.0}
\eea
with $\nu_{+}(0)=\nu_{-}(0)+1$ and $\nu_{-}(0) \sim 3$.

\subsection{Results for parton densities} \noindent

%{\bf The free parameters of parameterizations are fixed by a comparison with 
%  recent numerical solution of the DGLAP equation done by the NNPDF group\cite{NNPDF4}
%  and presented in Table 1.}

\begin{table}[h] 
\centering
\begin{tabular}{lccccccc}
	\hline
	\hline
	\\
	& $Q_0$, GeV & $A_V(0)$ & $A_q$ & $A_g$ & $B_V(0)$ & $B_{-}(0)$ & $B_{+}(0)$ \\
	\\ 
	\hline
	\\
	AKL &  $\sqrt{0.43}$ & $3.0$ & $0.95$ & $0.77$ & $100.0$ & $0.0$ & $13\cdot10^6$ \\	
	\\
	\hline
	\\
	& $m_{q,-}$ & $m_{q,+}$ & $m_{g,-}$ & $m_{g,+}$ & $\nu_{V}(0)$ & $\nu_{-}(0)$ & $\nu_{+}(0)$ \\
	\\ 
	\hline
	\\
	AKL & $2.0$ & $1.0$ & $1.0$ & $2.0$ & $4.0$ & $7.2$ & $8.2$ \\	
	\\
	\hline
	\hline
\end{tabular}
	\caption{The fitted values of various parameters involved in our analytical expressions for PDFs in a proton.}
	\label{table1}
\end{table}

\iffalse

\begin{table}[h] 
\centering
\begin{tabular}{lcccccccccc}
	\hline
	\hline
	\\
	& $Q_0$, GeV & $A_V(0)$ & $B_V(0)$ & $A_q$ & $B_{-}(0)$ & $A_g$ & $B_{+}(0)$ &
        %$\mu_{i}(0)$ &
        $\nu_{-}(0)$ & $\nu_{V}(0)$ \\
	\\ 
	\hline
	\\
	AKL &  $\sqrt{0.43}$ & $3.0$ & $100.0$ & $0.95$ & $0.0$ & $0.77$ & $13\cdot10^6$ &
        %$1.0$ &
        $7.2$ & $4.0$ \\	
	\\
	\hline
	\hline
\end{tabular}
	\caption{The fitted values of various parameters involved in our analytical expressions for PDFs in a proton.}
	\label{table1}
\end{table}
\fi

{}From numerical analysis we have $B_{-}(s)=0$, i.e. the large $x$ behavior is defined by valence quarks.
%}
Then the results for $f_{a,-}(x,Q^2)$ with $a=q$ or $g$ in~(\ref{q+l}) are strongly simplified:
\bea
%\z f_j(x,Q^2) = \sum_{\pm}\, f_{j,\pm}(x,Q^2),~~ (j=q,g)
%\label{sg_l} \\
\z f_{q,-}(x,Q^2)=
\biggl[A_{q}e^{- d_{-} s} (1-x)^{m_{q,-}} +
  %\frac{B_{-}(s)\, x}{\Gamma(1+\nu_{-}(s))}+
  D_{-}(s)x (1 -x)  \biggr]\,  (1-x)^{\nu_{-}(s)}, \nonumber \\
%\label{q-l.1}\\
\z f_{g,-}(x,Q^2)=
%\biggl[
A_{g}^{-}e^{- d_{-} s}
%(1-x )^{m_{g,-}} \nonumber\\
%&& + \frac{K_{-}}{\Gamma(2+\nu_{-}(s))}  \,
%  \frac{B_-(s) x}{\left[\ln(1/(1-x))+\hat{c} + \Psi(\nu_{-}+2)\right]} \biggr]  
\,  (1-x)^{\nu_{-}(s)+m_{g,-}+1} \,.
%\nonumber \\
%%\label{g-l.1}\\
%%\z f_{g,+}(x,Q^2)= 
%%\biggl[
%%A_{g}^+ \overline{I}_0(\sigma)e^{-\overline d_{+} s} (1-x)^{m_{g,+}} + \frac{B_{+}(s)\, x}{\Gamma(1+\nu_{+}(s))} + D_{+}(s)x (1 -x)  \biggr]\, (1-x)^{\nu_{+}(s)} \,, \label{g+l}\\
%\z f_{q,+}(x,Q^2)= 
%%\biggl[
%A_{q}^+ \tilde{I}_1(\sigma)e^{-\overline d_{+} s}
%  %(1-x)^{m_{q,+}}\nonumber \\
%%   && - \frac{ K_{+}}{\Gamma(2+\nu_{+}(s))}  \, \frac{B_+(s) x}{\left[\ln(1/(1-x))+\hat{c} + \Psi(\nu_{+}+2)\right]} \biggr] 
%  (1-x)^{\nu_{+}(s)+m_{q,+}+1}
%%  % + B_g(s)x^{\mu_{+}(s)}
  \label{q+l.1}
  \eea
%and, thus, the form (\ref{S2.3}) is exactly reconstructed.
\noindent
Similar simplification has the place also for $f_{a,-}(x,Q_0^2)$ in~(\ref{q+.0}). To have it,  we should put $s=0$ in the
%above
results (\ref{q+l.1}). 

Moreover, we have simplifications also for $G^{-}_a(s)$ with $a=q$ or $g$
in (\ref{G+s}), for $\overline{G}^{-}_q(s)$ in (\ref{oG+g_l}) and for $G^{-}_a(s=0)$ in (\ref{MC+.s0-l.1}). Indeed, we should replace  $G^{-}_a(s)$ 
in (\ref{G+s}) by
\bea
\z G^{-}_q(s) = \frac{A_{q}}{1+\nu_{-}(s)+m_{q,-}}\, e^{-d_{-}s}
%+\frac{B_{-}(s)}{\Gamma(3+\nu_{-}(s))}
+ D_{-}(s)\frac{\Gamma(3/2)\Gamma(2+\nu_{-}(s))}{\Gamma(7/2+\nu_{-}(s))}
\,, \nonumber \\
\z G^{-}_g(s) = \frac{A^{(-)}_{g}}{2+\nu_{-}(s)+m_{g,-}}\, e^{-d_{-}s}
\,.
\label{G+s.R}
\eea
and $G^{-}_a(s=0)$ $(a=q,g)$ in (\ref{MC+.s0-l.1}) by the results (\ref{G+s.R}) with $s=0$.
The results $\overline{G}^{-}_q(s)$ in (\ref{oG+g_l}) should be replaced by
\be
%\bea \z
\overline{G}^{-}_q(s) = \frac{A_{q}}{1+\nu_{-}(s)+m_{q,-}}\, e^{-d_{-}s}\,.
%+\frac{B_{-}(s)}{\Gamma(3+\nu_{-}(s))}\,, \nonumber \\
\label{oG-s_l}
\ee

The values of all parameters involved into 
derived expressions can be determined from the comparison with the
known parametrizations of numerical solutions of DGLAP equations 
and/or taking into account the sum rules. 
In our analysis, we employ the latest 
parametrizations proposed by the NNPDF Collaboration, namely, NNPDF4.0 set\cite{NNPDF4}.
Results of our fit are collected in Table~1.
Additionally, in Fig.~1 we show the comparison between our 
PDFs (labeled as AKL) and corresponding results obtained by the 
MMHT'2014\cite{MMHT} and NNPDF groups.
We find a good agreement between our analytical 
derivation and relevant numerical analyses. 

\begin{figure}
\begin{center}
\includegraphics[width=6cm]{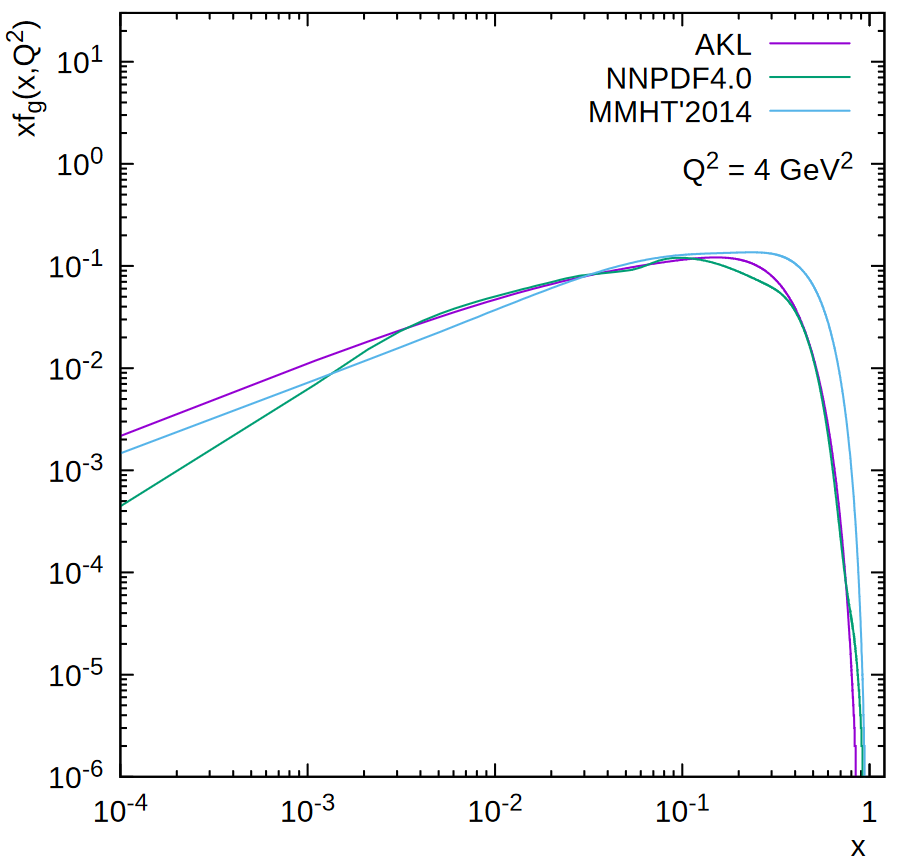}
\includegraphics[width=6cm]{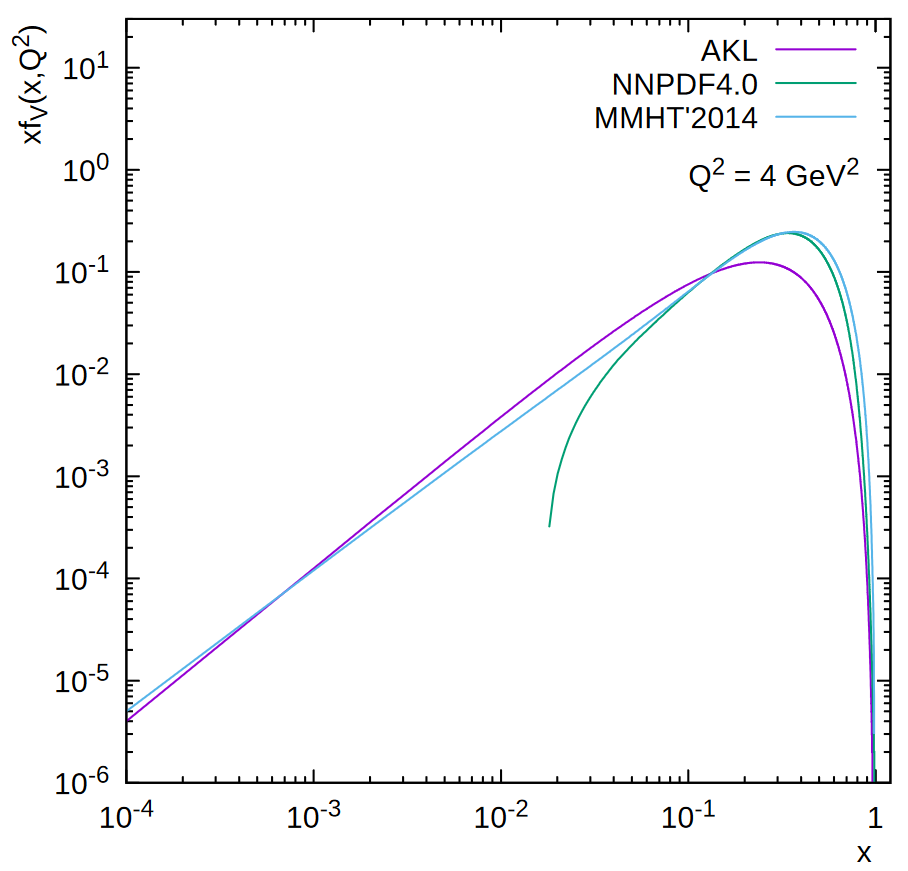}
\includegraphics[width=6cm]{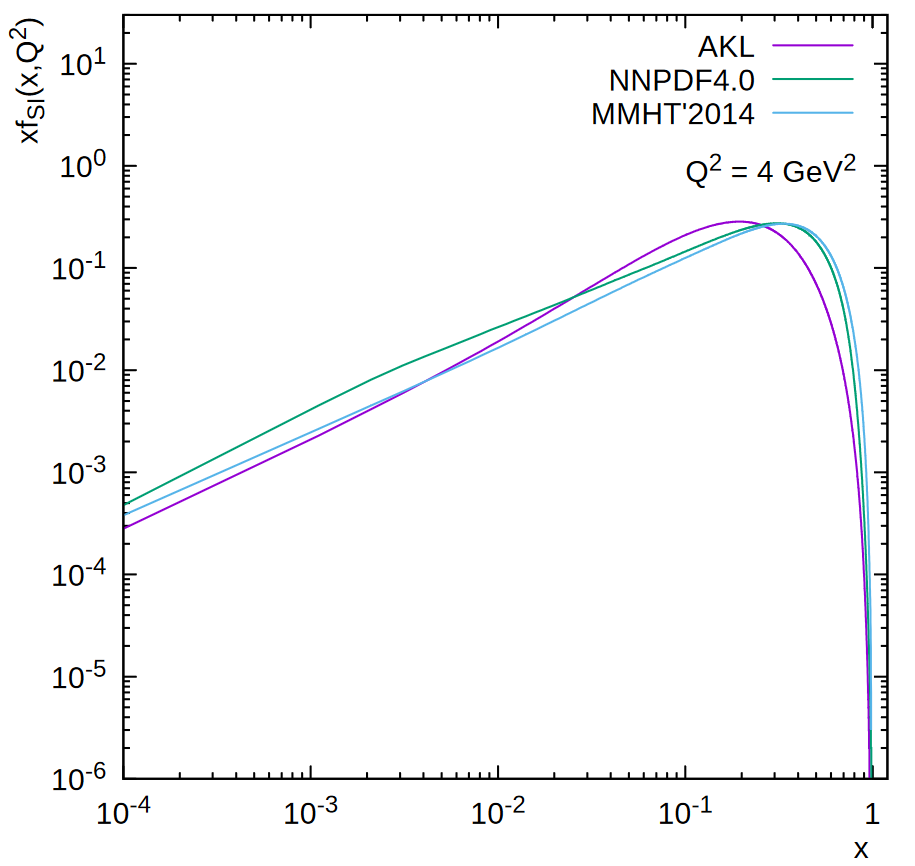}
\includegraphics[width=6cm]{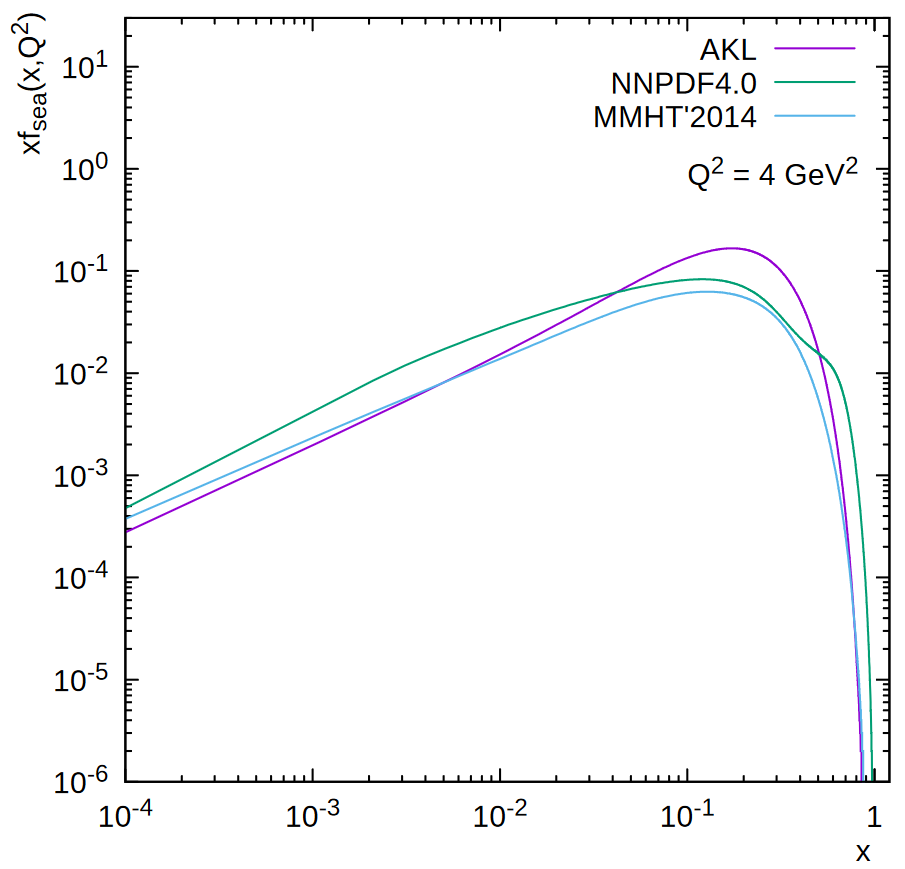}
\caption{The gluon, valence, singlet and sea quarks densities in a proton calculated as a function of the longitudinal momentum fraction $x$ 
at hard scale $Q^2=4$ GeV${}^2$. The purple curve corresponds to the results obtained with AKL,
the green and blue curves with NNPDF4.0 (LO) and MMHT'2014 (LO) parton density functions, respectively.}
\label{figpngs}
\end{center}
\end{figure}

%Singlet:
%x<0.05 chi2 = 0.46; x<0.1 chi2 = 0.61; x<0.5 chi2 = 5.66; x<1 chi2 = 10.32;
%Valence:
%x<0.05 chi2 = 0.91; x<0.1 chi2 = 1.32; x<0.5 chi2 = 8.06; x<1 chi2 = 11.7;
%Gluon:
%x<0.05 chi2 = 0.7; x<0.1 chi2 = 1.1; x<0.5 chi2 = 2.84; x<1 chi2 = 3.13;\\
%{\bf In Fig. 1 only first part (from three ones) is demonstrated}

\section{TMD parton densities in a proton} \indent

Now we turn to the derivation of analytical expressions for the 
TMD gluon and quark density functions in a proton. 
Our consideration is based on the KMR
%Kimber-Martin-Ryskin (KMR)
procedure\cite{21}, which is a formalism to construct the 
TMDs from the conventional PDFs.
The key assumption is that the transverse momentum dependence of the 
parton distributions enters only at the last evolution step,
so that conventional PDFs can be used up to this step.
There are known differential and integral 
formulations of KMR approach 
in the literature (see\cite{Golec-Biernat:2018hqo} for more information and discussion). 
Below we derive expressions for the TMDs 
using both these schemes.

%\subsection{differential approach} \indent
\subsection{Differential formulation} \noindent

In the differential formulation of KMR procedure, we have the TMD parton densities 
$f_a^{(d)}(x,k^2,Q^2)$, where $a=V$, $q$ or $g$ as
\be
f^{(d)}_a(x,k^2,Q^2) = \frac{\partial}{\partial \ln k^2} \Bigl[T_a(Q^2,k^2) \hat{D}_a(x,k^2)\Bigr] \, ,
\label{Dif} 
\ee
where
\be f_a(x,k^2)=x\hat{D}_a(x,k^2)\,.
\label{Df} 
\ee
\noindent
Since
\be
f_a(x,k^2) =\sum_{\pm} f_{a,\pm}(x,k^2),~~a=q,g,
\label{fpm} 
\ee
we see that the $f_a^{(d)}(x,k^2,Q^2)$ have similar form
\be
f_a^{(d)}(x,k^2,Q^2) =\sum_{\pm} f_{a,\pm}^{(d)}(x,k^2,Q^2).
\label{un_fpm} 
\ee
\noindent
Using the expressions (\ref{S4.1_l}) --- (\ref{q+l}) for PDFs $f_a(x,k^2)$,
we obtain for the TMDs.
The complete results are shown in Appendix C. Here we present only the final results:
%{\bf with
%  %{\bf We see that results %(\ref{Unfq-}),  (\ref{Unfg-}) simplify if
%  $B_{-}(s)=0$:}
\bea
\z f^{(d)}_{i}(x,k^2,Q^2) = \beta_0 \, a_s(k^2) T_q(Q^2,k^2) \times \Biggl\{ \left[d_q \, R_q(\Delta)-r_{V} \ln\left(\frac{1}{1-x}\right)\right] 
    \, f_{i}(x,k^2) \nonumber \\
    \z - \Biggl[d_{V}(1-\lambda_V) \,A_V\; e^{-d_V(1-\lambda) s_2} (1-x) \nonumber \\
      &&+\left[p_V+r_{V}\Psi(1+\nu_V(s_2))\right] \frac{B_{i}(s_2)\, x}{\Gamma(1+\nu_{V}(s_2))}
      %+ \stackrel{*}{D_V}(s_2) x(1-x)
      \Biggr]
      (1-x)^{\nu_{V}(s_2)} \Biggr\},~~(i=V,NS)\,, \nonumber \\
%\label{UnfV}\\
\z f^{(d)}_{g,+}(x,k^2,Q^2) = \beta_0 \, a_s(k^2) T_g(Q^2,k^2) \times \Biggl\{ \left[d_g \, R_g(\Delta)-r_{+} \ln\left(\frac{1}{1-x}\right)\right] 
    \, f_{g,+}(x,k^2) \nonumber \\
    \z -\Biggl[\left[\hat{d}_+ \overline{I}_1(\sigma_2)+ \overline d_{+} \overline{I}_0(\sigma_2)
        %\frac{\hat{d}_+}{\rho_2} I_1(\sigma_2) + \overline d_{+} I_0(\sigma_2)
  %+ d_{-} f_a^-(x,k^2)
        \right] \,A_g^{+}\; e^{-\overline d_{+} s_2} (1-x)^{m_{g+}}
      \nonumber \\
      &&+\left[p_+ + r_{+}\Psi(1+\nu_+(s_2))\right] \frac{B_{+}(s_2)\, x}{\Gamma(1+\nu_{+}(s_2))}
      %+ \stackrel{*}{D^+_g}(s_2) x(1-x)
      \Biggr]
     (1-x)^{\nu_{+}(s_2)} \Biggr\}
\, , \nonumber \\
%\label{Unfg+}\\
  \z f^{(d)}_{q,+}(x,k^2,Q^2) = \beta_0 \, a_s(k^2) T_q(Q^2,k^2) \times \Biggl\{ \left[ d_q \, R_q(\Delta)-r_{+} \ln\left(\frac{1}{1-x}\right)\right] 
    \, f_{q,+}(x,k^2) \nonumber \\
    \z -\left[\hat{d}_+\, \overline{I}_0(\sigma_2) + \overline d_{+} \tilde{I}_1(\sigma_2)
      %\rho_2 I_1(\sigma_2)
  %+ d_{-} f_a^-(x,k^2)
  \right] \,A_q^{+}\; e^{-\overline d_{+} s_2} (1-x)^{\nu_{+}(s_2)+1+m_{q+}} \Biggr\}\, , \nonumber \\
%\label{Unfq+}\\
\z f^{(d)}_{q,-}(x,k^2,Q^2) = \beta_0 \, a_s(k^2) T_q(Q^2,k^2) \times \Biggl\{ \left[d_q \, R_q(\Delta)-r_{-} \ln\left(\frac{1}{1-x}\right)\right] 
    \, f_{q,-}(x,k^2) \nonumber \\
    \z -
    %\Biggl[
    d_{-} \,A_q\; e^{-d_{-} s_2}
    %(1-x)^{m_{q-}} \nonumber \\
    %  &&+\left[p_-+r_{-}\Psi(1+\nu_-(s))\right] \frac{B_{-}(s)\, x}{\Gamma(1+\nu_{-}(s))}+ \stackrel{*}{D^-_q}(s)  x(1-x)\Biggr]
      (1-x)^{\nu_{-}(s_2)+m_{q-}} \Biggr\}
\, , \nonumber \\
%\label{Unfq-.1}\\
\z f^{(d)}_{g,-}(x,k^2,Q^2) = \beta_0 \, a_s(k^2) T_g(Q^2,k^2) \times \Biggl\{ \left[d_g \, R_g(\Delta)-r_{-} \ln\left(\frac{1}{1-x}\right)\right] 
    \, f_{g,-}(x,k^2) \nonumber \\
    \z -d_{-} \,A_g^{-}\; e^{-d_{-} s_2}
    %(1-x)^{m_{g-}}
    %%\nonumber \\&&
    %-\left[p_-+r_{-}\Psi(1+\nu_-(s))\right] \nonumber \\&& \times
 %\frac{K_{-}}{\Gamma(2+\nu_{-}(s))}  \,
 % \frac{B_-(s) x}{\left[\ln(1/(1-x))+\hat{c} + \Psi(\nu_{-}+2)\right]} 
  \,  (1-x)^{\nu_{-}(s_2)+m_{g-}+1} \Biggr\}
\, ,
\label{Unfg-.1}
\eea
where we neglected the derivations of $D_{a}(s)$ $(a=V,\pm)$.
%, the magnitudes of intermediate terms.
Indeed, the magnitudes $D_{a}(s)$ of intermediate terms 
have slow $s$ dependence and their derivations
can be neglected.

\subsection{
  %TMDs from
  Integral formulation
%  of KMR approach
} \noindent
%{\bf!!! Here we use the condition $B_{-}(s)=0$ and obtain the following results:}

Following investigations done in\cite{Kotikov:2019kci}, we can obtain the following results ($j=V$, $NS$):
\bea
\z f^{(i)}_{j}(x,k^2,Q^2) = \beta_0 a_s(k^2) \, T_q(Q^2,k^2)\, 
\times \Biggl\{ \left[d_q \, R_q(\Delta)-r_{V} \ln\left(\frac{1}{1-x}\right)\right]\, f_{j}(x,k^2) -
\tilde{R}_{j}(x,k^2) \Biggl\},~~
%(j=V,NS),
\nonumber \\
&& f^{(i)}_{a}(x,k^2,Q^2) = f^{(i)}_{a,+}(x,k^2,Q^2) + f^{(i)}_{a,-}(x,k^2,Q^2) ,~~ \nonumber \\
\z f^{(i)}_{a,-}(x,k^2,Q^2) =\beta_0 a_s(k^2) \, T_a(Q^2,k^2)\,
\times \Biggl\{ \left[d_a \, R_a(\Delta)-r_{-} \ln\left(\frac{1}{1-x}\right)\right]\,\, f_{a,-}(x,k^2) 
- \tilde{R}_{a,-}(x,k^2)\Biggl\},~~ \nonumber \\
\z f^{(i)}_{a,+}(x,k^2,Q^2) =\beta_0 a_s(k^2) \, T_a(Q^2,k^2)\,
\times \Biggl\{ \left[d_a \, R_a(\Delta)-r_{+} \ln\left(\frac{1}{1-x}\right)\right]\,\, \overline{f}_{a,+}(x,k^2) 
- \tilde{R}_{a,+}(x,k^2)\Biggl\}
\label{Unfi}
\eea
\noindent
where
\bea
\z  \tilde{R}_{j}(x,k^2)=\Biggl[d_{V}(1-\lambda_V) \,A_j\; e^{-d_{V}(1-\lambda) s_2}\, \overline{x}^{\lambda}
  \nonumber \\&&
  +\left[p_V+r_{V}\Psi(1+\nu_V(s_2))\right] \frac{B_{j}(s_2)\, x}{\Gamma(1+\nu_{V}(s_2))}
  %\nonumber \\ && + r_V \,\frac{B_{V}(s_2)\, x}{\Gamma(1+\nu_{V}(s_2))} \left(\ln\left(\frac{\Delta}{1-x}\right)
  %+ S_1(\nu_V(s_2))\right)
  \Biggr]\, (1-x)^{\nu_{V}(s_2)+1}
%,~~(j=V,NS)
\,,  \nonumber \\
%\label{UnfiV}\\
\z  \tilde{R}_{q,+}(x,k^2)=
%\Biggl[
\Bigl(\hat{d}_{+}\, \overline{I}_0(\overline{\sigma}_2) + \overline{d}_{+}\, \tilde{I}_1(\overline{\sigma}_2)
  %\overline{\rho}_2 \, I_1(\overline{\sigma}_2)
  \Bigr)
 % (1-x)^{+m_{q,+}}
  %\nonumber \\
% && + r_{+} x\, \left(\ln\left(\frac{\Delta}{1-x}\right)
  %+ S_1(\nu_{-}(s_2)+m_{q,+}+2)\right)\,x
  %\Biggr]
  \,A_q^{+}\; e^{-\overline{d}_{+} s_2}
  %(1-x)^{+m_{q,+}}  + ... \Biggr]
  (1-x)^{\nu_{+}(s_2)+m_{q,+}+2}\,,  \nonumber \\
%\label{Unfiqpl}\\
%(1-x)^{\nu_{+}(s_2)+m_{q,+}+1}\,,  \label{Unfiqpl}\\
\z  \tilde{R}_{g,+}(x,k^2)=\Biggl[\Bigl(\hat{d}_{+}\, \overline{I}_1(\overline{\sigma}_2)
  %\frac{1}{\overline{\rho}_2}\, I_1(\overline{\sigma}_2)
  + \overline{d}_{+}\, \overline{I}_0(\overline{\sigma}_2)\Bigr) \,A_g^{+}\; e^{-\overline{d}_{+} s_2}\,
  (1-x)^{m_{g,+}} 
  \nonumber \\
  && +\left[p_++r_{+}\Psi(1+\nu_V(s_2))\right] \,\frac{B_{+}(s_2)\, x}{\Gamma(1+\nu_{+}(s_2))}
  %\, \left(\ln\left(\frac{\Delta}{1-x}\right)
  % + S_1(\nu_{+}(s_2))\right)
  \Biggr] 
  %(1-x)^{\nu_{+}(s_2)}  \,,  \label{Unfigpl}\\
(1-x)^{\nu_{+}(s_2)}  \,, \nonumber \\
%\label{Unfigpl}\\
\z  \tilde{R}_{q,-}(x,k^2)=
%\Biggl[
d_{-}\,A_q\; e^{-d_{-} s_2}
  %(1-x)^{m_{g,-}-1}
%  + ...
  %+ r_{-} \,D_{-}(s_2)\, x\, \left(\ln\left(\frac{\Delta}{1-x}\right)
  %+ S_1(\nu_{-}(s_2)+1)\right)
%  \Biggr]
%\nonumber \\&&\times
(1-x)^{\nu_{-}(s_2)+m_{g,-}}\,, \nonumber \\
%\label{Unfiqmi}\\
\z  \tilde{R}_{g,-}(x,k^2)=
%\Biggl[
d_{-}
    %(1-x) + r_{-} \,x\, \left(\ln\left(\frac{\Delta}{1-x}\right)
    %+ S_1(\nu_{-}(s_2)+m_{g,-}+1)\right)\Biggr]
\,A_g^{-}\; e^{-d_{-} s_2}
%+ ... \Biggr]
    %\nonumber \\
%&& \times \, (1-x)^{\nu_{-}(s_2)+m_{g,-}+1}\,,  \label{Unfigmi}
(1-x)^{\nu_{-}(s_2)+m_{g,-}+2}\,,  \label{Unfigmi}
    \eea
\noindent    
with
\be
%S_1(\nu)=\Psi(\nu+1)+\gamma_{E},~~
\overline{x}=\frac{x}{x_0},~~x_0=1-\Delta,~~\overline{\sigma}_2=\sigma_2(x\to \overline{x}),~~\overline{\rho}_2=\rho_2(x\to \overline{x})\,.
 \label{over.x}
\ee
\noindent
and
\be
\overline{f}_{a,+}(x,k^2)=f_{a,+}(x,k^2) ~~\mbox{with}~~ \sigma_2 \to \overline{\sigma}_2,~~\rho_2 \to \overline{\rho}_2\,.
 \label{over.f}
\ee
\noindent
As in~(\ref{Unfg-.1}), we neglected contributions coming from the magnitudes  $D_{a}(s)$, where $a=V$, $\pm$.

%The simbol $+ ...$ marks terms $\sim B_{..}$ and $\sim D_{..}$, which can be added later, if necessary.

\subsection{Sudakov form factors $T_a(Q^2, k^2)$} \noindent

%Evaluating~(\ref{Ta}),
For the Sudakov form factors, we have\cite{Kotikov:2021yzb}:
\be
T_a(Q^2,k^2) = \exp \Bigl[ -d_a R_a(\Delta) s_1 \Bigr] \, ,
\label{Ta.1}
\ee
where
\bea
&&s_1=\ln \left( \frac{a_s(k^2)}{a_s(Q^2)} \right),~ d_a=\frac{4C_a}{\beta_0},~ C_q=C_F,~ C_g=C_A,~
%d_g=\frac{4C_A}{\beta_0},~
\beta_0=\frac{C_A}{3} \Bigl(11-2\varphi\Bigr),~
\nonumber \\
&&R_q(\Delta)= \ln\left(\frac{1}{\Delta}\right) - \frac{3x_0^2}{4}
= \ln\left(\frac{1}{\Delta}\right) - \frac{3}{4} (1-\Delta)^2,~~ \nonumber \\
&&R_g(\Delta)= \ln\left(\frac{1}{\Delta}\right) - \left(1-\frac{\varphi}{4}\right) x_0^2
+ \frac{1-\varphi}{12} x_0^3 (4-3x_0) = \nonumber \\
&&= \ln\left(\frac{1}{\Delta}\right) - \left(1-\frac{\varphi}{4}\right) (1-\Delta)^2
+ \frac{1-\varphi}{12} (1-\Delta)^3 (1+3\Delta) \, .
\label{Ta.2}
\eea

\subsection{Cut-off parameter $\Delta$} \noindent

For the phenomenological applications, the cut-off parameter $\Delta$ usually has one of two basic forms:
\be
\Delta_1=\frac{k}{Q},~~\Delta_2=\frac{k}{k+Q},
\label{Delta12}
\ee

\noindent
that reflects the two cases: $\Delta_1$ is in the strong ordering, $\Delta_2$ is in the angular ordering (see \cite{Golec-Biernat:2018hqo}).
In all above cases, except the results for $T_a(Q^2,k^2)$, we can simply replace the parameter $\Delta$
by
%the ones
$\Delta_1$ and/or $\Delta_2$. For the Sudakov form factors, we note
that the parameters $\Delta_i$ (with $i=1, 2$) contribute to the integrand
%subintegral expressions
%in (\ref{Ta})
and, thus, their
momentum dependence changes the results in (\ref{Ta.1}). Perform the correct evaluation (see \cite{Kotikov:2021yzb})
%of the integral (\ref{Ta}),
%we should
%%it is simple to
%recalculate the $p^2$ integration in (\ref{Ta}). So,
we have
\be
T_a^{(i)}(Q^2,k^2) = \exp \left[ -4C_a \int\limits^{Q^2}_{k^2} \, \frac{dp^2}{p^2} \, a_s(p^2) R_a(\Delta_i)  \right].
\label{Ta.2}
\ee

\noindent
The analytic evaluation of $T_a^{(i)}(Q^2,k^2)$ is a very cumbersome procedure,
%and
which will be
accomplished in the future. With the purpose of simplifying
our analysis, below we use the numerical
results for $T_a^{(i)}(Q^2,k^2)$.

\subsection{Results for TMD parton densities} \indent

\begin{figure}
\begin{center}
\includegraphics[width=6cm]{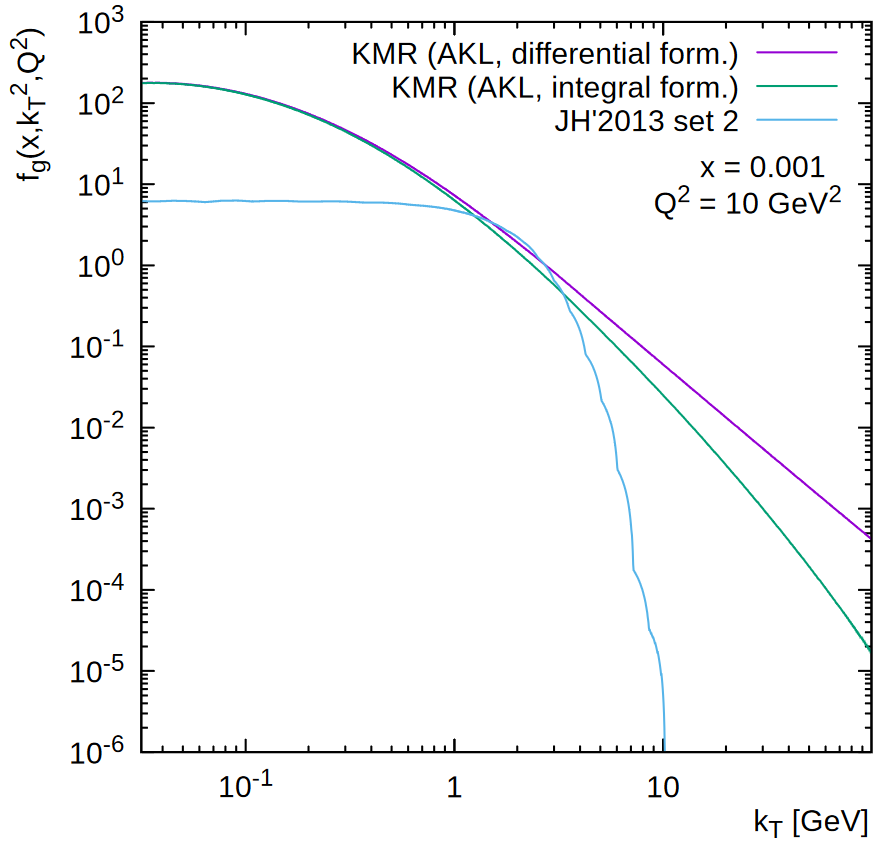}
\includegraphics[width=6cm]{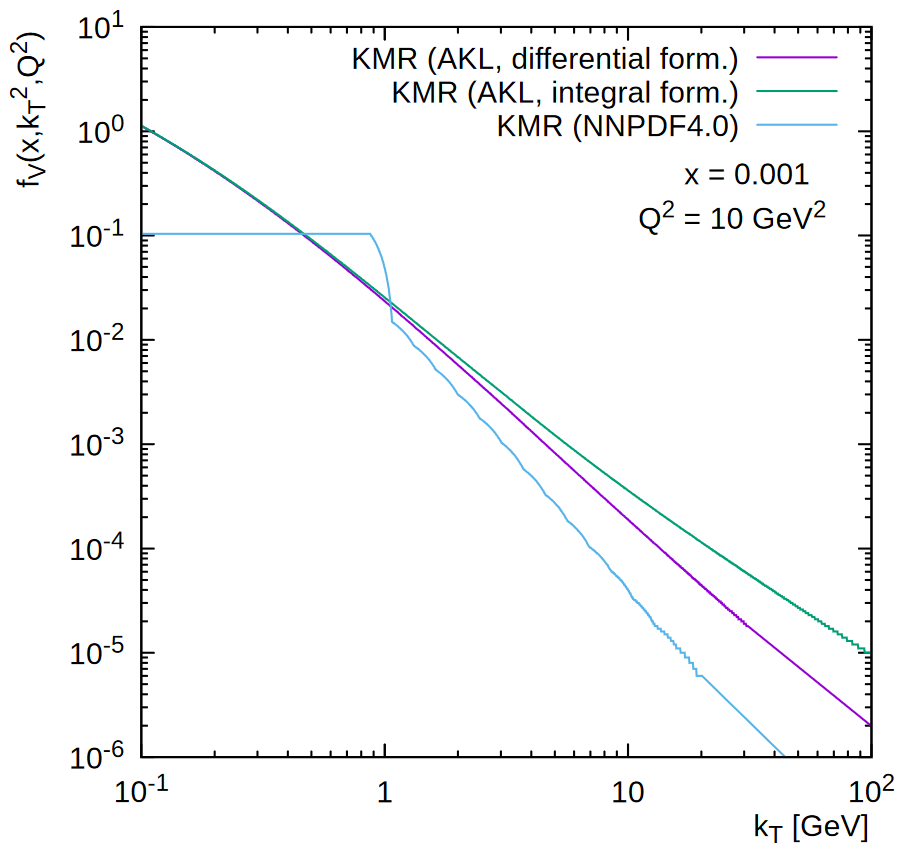}
\includegraphics[width=6cm]{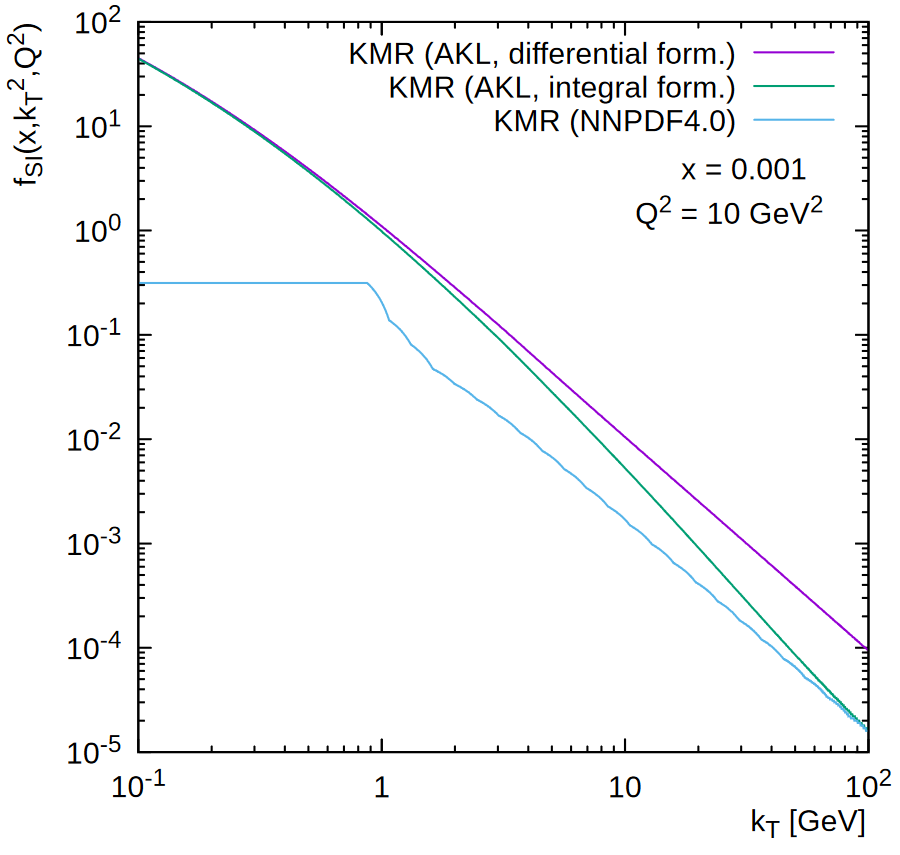}
\includegraphics[width=6cm]{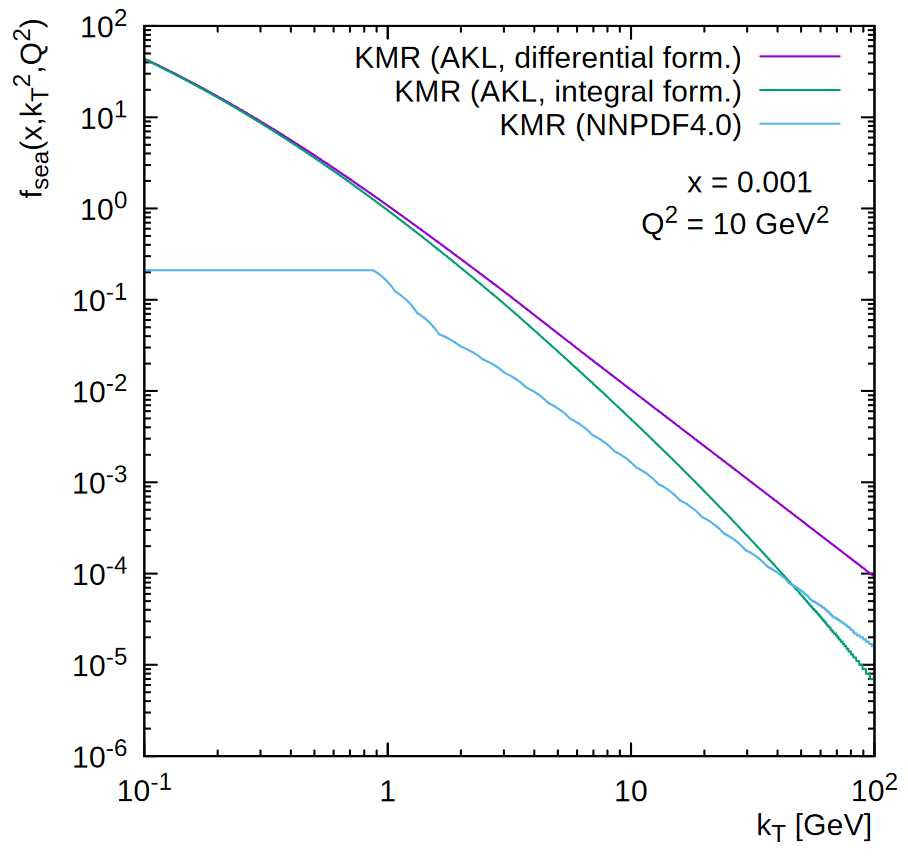}
\caption{The TMD gluon, valence, singlet and sea quarks densities in a proton calculated as a function of the parton
transverse momentum $k_T^2$ at longitudinal momentum fraction $x = 0.001$ and
hard scale $Q^2 = 10$ GeV${}^2$. The purple and green curves correspond to the results obtained with AKL sets, 
which were obtained from differential and integral formulation of KMR approach, respectively.
The blue curve corresponds to the results obtained with TMD gluon distribution JH'2013 set 2 and with TMD quark distributions 
calculated numerically in the traditional KMR scenario,
where the conventional parton densities from standard NNPDF (LO) set are used as an input.}
\label{figTMD}
\end{center}
\end{figure}

Our results for the TMD quark and gluon densities in a proton,
obtained in both differential and integral formulation of the 
KMR procedure, are shown in Fig.~2.
Note that cut-off parameter $\Delta$ is taken in the form 
corresponding to the angular ordering condition.
We find that the difference 
between these two scenarios is originated at large parton transverse momenta only,
whereas at low $k_T$ they are coincide to each other.
In addition, we plot here for comparison the 
results derived in other approaches.
So, we show the TMD gluon distribution 
obtained from the numerical solution\cite{JH} of the 
Catani-Ciafaloni-Fiorani-Marchesini (CCFM) evolution equation\cite{CCFM},
namely, JH'2013 set 2. The CCFM equation 
is smoothly interpolates between the small-$x$ BFKL gluon 
dynamics and large-$x$ DGLAP one, and 
JH'2013 set 2 gluon is often used in a different phenomenological
applications (see, for example,\cite{kt-app1,kt-app2,kt-app4,kt-app5}).
Also we plot here the resuls for the TMD PDFs 
obtained within the KMR approach where the NNPDF4.0 
set has been used as an input.
One can see that all these TMDs have
a different shape in $k_T$
and a different overall normalization.
The step behavior at low $k_T \sim 1$~GeV
is related to the special normalization 
condition usually applied in the KMR scheme (see\cite{21} for more information).
Studying the phenomenological consequences
of observed differences is an important and interesting task,
but, however, it is out of our present consideration.

\section{Conclusion} \indent

In the paper we proposed an analytical expressions
for the proton PDFs based on their exact asymptotics at small and large $x$ values.
The derived parameterizations contain subasymptotic terms, which are fixed by 
momentum conservation and, in the nonsinglet and valence parts, by the Gross-Llewellyn-Smith
and Gottfried  sum rules (see\cite{GrLl} and \cite{Gottfried:1967kk}, respectively. 
The rest of parameters is fixed by a comparison with 
  recent numerical solution of the DGLAP equation done by the NNPDF group\cite{NNPDF4}
  and presented in Table 1.
Then, our consideration has been extended to the parton densities,
dependent on the transverse momentum (TMDs).
These quantities are often used in the number of phenomenological 
applications and widely discussed in the literature at present.
We employ the popular Kimber-Martin-Ryskin formalism \cite{21,22}
and derive the TMD quark and gluon distributions in a proton
within the both differential and integral formulation of the KMR scheme.
In the calculations we considered the different treatments of kinematical constraint, 
reflecting the angular and strong ordering conditions.
An analytical expressions for the PDFs and TMDs (in particular, for the quark TMDs), 
valid at both low and large $x$, were obtained for the first time.
As a next step, we plan to use the present PDF sets to study PDF modifications in nuclei 
and, of course, to extend the present investigations beyond the LO.
In these investigations we will follow\cite{Kotikov:1988aa,62} and \cite{72}, respectively.

\section*{Acknowledgements} \indent

We thank M.A.~Malyshev for very useful discussions and remarks.
Researches described in Sections~3 --- 5 were 
supported by the Russian Science Foundation under grant 22-22-00387.
Studies described and performed in Section 6
were supported by the Russian Science Foundation under grant 22-22-00119.

\appendix
\def\theequation{A\arabic{equation}}
\setcounter{equation}{0}

\section{ PDF asymptotics at large $x$ values}

Using large $n$ expansion for $S_1(n)$ as
\be
S_1(n) = \ln n + \gamma_E +O(n^{-1})\, ,
\label{S1n}
\ee
of some auxiliary anomalous dimension $d_c=c_1 S_1(n)+ c_2$,
we see for the general renorm-group exponent
\be
e^{-d_cs}= e^{-(c_1(ln n+\gamma_E)-c_2)s} +O(n^{-1})= n^{-c_1 s} e^{-\overline{c}_2 ss} +O(n^{-1}),~~ \overline{c}_2= c_2+c_1 \gamma_E \,.
\label{RGExp}
\ee

Let some PDF $f_c(x,Q^2)$ has the following $Q^2$-dependence
\be
f_c(x,Q^2)= f_c(x,Q_0^2)\,  e^{-d_cs} = f_c(x,Q_0^2)\,  n^{-c_1 s} e^{-\overline{c}_2 s} \,.
\label{fc}
\ee

It is convenient to represent our basic variable $s$ shown in Eq. (\ref{d_a}) as
\be
s=\overline{s}(Q^2)-\overline{s}(Q_0^2) \equiv \overline{s}-\overline{s}_0\,,
\label{sOs}
\ee
when
\be
\overline{s}(Q^2) = \ln[\ln(Q^2/\Lambda_{\rm LO}^2)] \,.
\label{Os}
\ee

When we can see that the result (\ref{fc}) for PDF $f_c(x,Q^2)$ can be rewritten
%represented
as
\be
f_c(x,Q^2)= B_c \,  n^{-\nu_1-c_1 \overline{s}} e^{\nu_2-\overline{c}_2 \overline{s}} \,,
\label{fc.1}
\ee
with some free parameters $B_c$, $\nu_1$ and $\nu_2$. For the initial condition $f_c(x,Q^2_0)$, we have the same result but with the replace $s \to s_0$. 

Now we consider
%try to represent
the Mellin moment
\be
M_c(n) =\int_0^1 \, dx \, x^{n-1} \, (1-x)^{\nu_c} = \frac{\Gamma(n)\Gamma(\nu_c+1)}{\Gamma(n+\nu_c+1)}
\label{Mc}
\ee
and try to represent it  in a form, similar to (\ref{fc}) and (\ref{fc.1}).

It is convenient to use Sterling formula
%, which has the following form at large $z$ values
\be
\ln \Gamma(z) = z\ln z - z +\frac{1}{2} \ln \frac{2\pi}{z} + \frac{B_2}{2z} + O(z^{-2})\,
\label{Sti}
\ee
 at large $z$ values,
where $B_2$ is Bernoulli number.

So, for large $z$ and fixed $\delta$ values, we have after little algebra
\be
\ln \frac{\Gamma(z+\delta)}{\Gamma(z)} = \delta\ln z +\frac{\delta(\delta-1)}{2z} + O(z^{-2})
\label{Sti.1}
\ee
and, thus,
\be
\frac{\Gamma(z)}{\Gamma(z+\delta)} = z^{-\delta}e^{\delta(1-\delta)/(2z)}  + O(z^{-2}) = z^{-\delta} \left(1+\frac{\delta(1-\delta)}{2z}\right)  + O(z^{-2})
\label{Sti.2}
\ee

So, the Mellin moment $M_c(n)$ can be represented as
\be
M_c(n) = n^{-(\nu_c+1)} \left(1-\frac{\nu_c(1+\nu_c)}{2n}\right)  + O(n^{-2})
\label{Mc.1}
\ee

For PDF we have ($(\alpha+1) \to \nu_1+c_1\overline{s}$)
\be
f_c(x,Q^2)=B_c \frac{(1-x)^{\nu_c+ c_1\overline{s}}}{\Gamma(\nu_c+1+ c_1\overline{s})} e^{\nu_2 -\overline{c}_2\overline{s}} = f_c(x,Q_0^2)
\,\frac{\Gamma(\nu_c+1+ c_1\overline{s}_0)}{\Gamma(\nu_c+1+ c_1\overline{s})} e^{-\overline{c}_2s},
\label{fc.1}
\ee
where $\nu_c=\nu_1-1$. The constant $\nu_2$ may be neglected.

So, finally we have
\be
f_c(x,Q^2)= \overline{B}_c \frac{(1-x)^{\nu_c(s)}}{\Gamma(\nu_c(s)+1)}\, e^{-\overline{c}_2s},~~\nu_c(s)=\nu_c(0) +\overline{c}_2s\,,
\label{fc.1}
\ee
where
\be
\nu_c(0)=\nu_c+ c_1\overline{s}_0,~~\overline{B}_c = f_c(x,Q^2) \,  \frac{\Gamma(\nu_c(0)+1)}{(1-x)^{\nu_c(0)}}\, e^{\overline{c}_2\overline{s}_0} 
\label{oBc}
\ee
  
\subsection{$O(n^0)$ accuracy}

At $O(n^0)$ accuracy,
\be
\gamma_{qg}=\gamma_{gq}=O(n^{-1})\,,
\label{Gqg}
\ee
and, thus, the $Q^2$-evolutions of the singlet quark and gluon densities are not related each other.

The corresponding anomalous dimesions have the following form
\be
\gamma_a=8C_F\left(\ln n + \gamma_E -\frac{3}{4}\right)+ O(n^{-1}),~~(a=NS,V,qq), \gamma_a=8C_A\left(\ln n +\gamma_E\right)-2\beta_0+ O(n^{-1}),
\label{Gqg}
\ee
and parton densities
\be
f_q(x,Q^2)=f_{-}(x,Q^2)+ O(n^{-1}),~~f_g(x,Q^2)=f_{+}(x,Q^2)+ O(n^{-1}),
\label{fqg}
\ee
have the large $x$ asymptotics
\be
f_b(x,Q^2)= B_b(s) \frac{(1-x)^{\nu_b(s)}}{\Gamma(\nu_b(s)+1)},~B_b(s)=B_b(0) e^{-p_bs},~~
\nu_b(s)=\nu_b(0) +r_bs\,,~~(b=NS,V,\pm)
\label{fb}
\ee
where where $r_{b}$ and $p_{b}$ are given in Eqs. (\ref{S2.4}).
%\be
%r_a=\frac{4C_F}{\beta_0},~~p_a= r_a \bigl(4\gamma_E+\hat{c}_a),~~(a=NS,V,-),\nonumber \\
%r_g=\frac{4C_A}{\beta_0},~~p_g= r_g \bigl(\gamma_E-\hat{c}_+\bigr)
%\label{cb}
%\ee
%and $\hat{c}_a$ and $\hat{c}_+$ can be found in Eq. (\ref{Cj)

\subsection{$O(n^{-1})$ accuracy}

At $O(n^{-1})$ accuracy,
\bea
&&\gamma_{qg}=-\frac{4f}{n} + O(n^{-2}), \gamma_{gq}=-\frac{4C_F}{n} + O(n^{-2}),\nonumber \\
&&\gamma_{qq}=8C_F\left(\ln n + \hat{c}_{-} +\frac{1}{2n}\right)+ O(n^{-2}),~~\gamma_{gg}=8C_A\left(\ln n + \hat{c}_{+} +\frac{1}{2n}\right)+ O(n^{-2})\,.~~
\label{Gqg.1}
\eea
where $\hat{c}_{\pm}$ are given in Eq. (\ref{Cj}).
%where
%\be
%b_q=\gamma_E -\frac{3}{4},~~b_g=\gamma_E -\frac{\beta_0}{4C_A}\,.
%\label{bqg}
%\ee

Using these results, we have
\be
\gamma_{+}=\gamma_{gg}+ O(n^{-2}),~~\gamma_{-}=\gamma_{qq} + O(n^{-2}),
\label{ga+}
\ee
i.e. the evolution of the ``$\pm$''-components are same as in the previous subsection.

The corresponding projectors have the following from:
\be
\alpha=1+ O(n^{-2}), \beta=\frac{K_{-}}{n}
%\frac{f}{2(C_A-C_F)n}
\, \frac{1}{\ln n+\hat{c}},~~ \tilde{\ep}=\frac{K_{+}}{n}\, \frac{1}{\ln n+\hat{c}},
\label{bqg}
\ee
where $K_{\pm}$ and $\hat{c}$ are given in Eq. (\ref{Cj}).
%\be
%\hat{b}=\gamma_E+ \frac{C_Ab_g-C_Fb_q}{C_A-C_F} = \gamma_E+ \frac{11C_A-2f-2CF}{12(C_A-C_F)}.
%\label{hb}
%\ee

Thus, now the contributions of initial quarks (gluons) to the final quarks (gluons) during $Q^2$ evolution 
are same as in the previous subsection (see Eq. (\ref{fb}))
but there are additional contributions:  initial quarks to gluons and vice versa. The last contributions
have unusual form $\sim  \frac{1}{n\ln n}$ at the large $n$ values.

So, for  the $Q^2$-evolutions of the singlet quark and gluon densities we have
\be
f_a(n,Q^2)= \hat{f}_a^+(n,Q_0^2)\, e^{-d_{+}s}+ \hat{f}_a^-(n,Q_0^2)\, e^{-d_{-}s}\,,
\label{faQ2}
\ee
where
\bea
&&\hat{f}_q^+(n,Q_0^2)=-\beta \, f_g(n,Q_0^2),~~ \hat{f}_q^-(n,Q_0^2)=f_q(n,Q_0^2)+\beta \, f_g^+(n,Q_0^2), \nonumber \\
&&\hat{f}_g^-(n,Q_0^2)= \ep\, f_q(n,Q_0^2),~~ \hat{f}_g^+(n,Q_0^2)=f_g(n,Q_0^2)-\ep\, f_q^+(n,Q_0^2),
\label{faQ0}
\eea

Thus, the NS and valence parton densities have the large $x$ asymptotics, shown in Eq. (\ref{S2.3}) of main text.
%\be
%f_j(x,Q^2)= B_j(s) \frac{(1-x)^{\nu_j(s)}}{\Gamma(\nu_V(s)+1)},~B_j(s)=B_j(0) e^{-\overline{c}_Vs},~~
%\nu_V(s)=\nu_V(0) +c_qs\,,~~(j=NS,V)\,,
%\label{fj}
%\ee
%which is obtained with neglection the terms $\sim s/n$ during the $Q^2$-evolution. These results are exactly same as it was in the previous subsection.
%
For the singlet quark and gluon densities the situation is different. Their large $x$ asymptotics contain standard contributions: the``$+$''-component for gluon density
and the``$-$''-component for sea quark density (see Eq. (\ref{fb}), for example). But there are also
%Their $Q^2$-evolutions contain also the
%results (\ref{fj}), obtained in the previous subsection, but there are
the additional parts, $\sim K_{\pm}$ in Eq. (\ref{S2.3}), coming from the contributions $\sim \beta $ and $\sim \ep$ in (\ref{faQ0}).  
%\bea
%&&f_q(x,Q^2)= f_{-}(x,Q^2) + \overline{f}_{+}(x,Q^2),~~ f_g(x,Q^2)= f_{+}(x,Q^2) + \overline{f}_{-}(x,Q^2), \nonumber \\
%&&f_{\pm}(x,Q^2)= B_{\pm}(s) \frac{(1-x)^{\nu_{\pm}(s)}}{\Gamma(\nu_{\pm}(s)+1)},~B_{\pm}(s)=B_{\pm}(0) e^{-\overline{c}_{\pm}s},~~\\
%\label{fqg}
%\eea
We will sketch a way to calculate the additional parts $\overline{f}_{\pm}(x,Q^2)$ in Appendix B.

%\appendix
\def\theequation{B\arabic{equation}}
\setcounter{equation}{0}

\section{Results at large $\nu$ values}

To obtain the results $\overline{f}_{\pm}(x,Q^2)$ it is convenient to calculate
%firstly
the inverse Mellin transform of auxiliary function
\be
f_A(n)=\frac{1}{n^{\nu +2} (\ln n+ a)}\, .
\label{Af}
\ee

To do it, we expand the denominator of $f_A(n)$ as
\be
f_A(n)=\frac{1}{n^{\nu +2} (\ln n+ a)} = \sum_{m=0}^{\infty} \frac{(-1)^m \ln^m n}{n^{\nu +2} a^{m+1}} = \sum_{m=0}^{\infty} \frac{1}{a^{m +1}} {\left(\frac{d}{d\nu}\right)}^m\frac{1}{n^{\nu +2}}
\label{Af.exp}
\ee

Since
\be
\int_0^1 dx \, x^{n-2} \, (1-x)^{\nu+1} \approx \frac{\Gamma(\nu+2)}{(n-1)^{\nu+1}} \approx   \frac{\Gamma(\nu+2)}{n^{\nu+1}}\, ,
\label{Mellin}
\ee
then the function $f_A(x)$, which is the inverse Mellin transform of $f_A(n)$
\be
f_A(n)=\int_0^1 dx \, x^{n-2} \,f_A(x)\,,
\label{Afn}
\ee
has the form
\be
f_A(x)= \sum_{m=0}^{\infty} \frac{(-1)^m}{a^{m +1}} {\left(\frac{d}{d\nu}\right)}^m\frac{(1-x)^{\nu +1}}{\Gamma(\nu+2)}
=\sum_{m=0}^{\infty}  \frac{(-1)^m}{a^{m +1}} \sum_{k=0}^m \, C^m_k {\left(\frac{d}{d\nu}\right)}^{m-l}(1-x)^{\nu +1} {\left(\frac{d}{d\nu}\right)}^{l} \frac{1}{\Gamma(\nu+2)}\,.
\label{Afx}
\ee

In the r.h.s. we will have the powers of Polygamma functions
\be
\Psi(\nu+2)= \frac{d}{d\nu} \, \ln\Gamma(\nu+2),~~ \Psi^{(m+1)}(\nu+2)= \frac{d}{d\nu} \, \Psi^{(m)}(\nu+2),~~(m\geq 0)\,.
\label{Psi}
\ee

At large $\nu$-values, $\Psi(\nu+2) \sim \ln(\nu+2)$ and $\Psi^{(m)}(\nu+2) \sim 1/(\nu+2)^m$, $(m \geq 1)$. So, we can neglect contributions from  $\Psi^{(m)}(\nu+2)$, $(m \geq 1)$
%So, we have
and obtain
\be
{\left(\frac{d}{d\nu}\right)}^{l} \frac{1}{\Gamma(\nu+2)} \approx \frac{(-1)^l \Psi^l(\nu+2)}{\Gamma(\nu+2)}\,.
\label{Psi.1}
\ee

So, we have
\be
f_A(x) \approx \sum_{m=0}^{\infty} \frac{(-1)^m}{a^{m +1}} \left(\ln \frac{1}{1-x} + \Psi(\nu+2)\right)^m\, \frac{(1-x)^{\nu +1}}{\Gamma(\nu+2)}
=\frac{1}{\ln \frac{1}{1-x} +a+ \Psi(\nu+2)}\, \frac{(1-x)^{\nu +1}}{\Gamma(\nu+2)}\,.
\label{Afx.1}
\ee

To obtain a contribution $\sim K_{-}$ in Eqs. (\ref{G+s}) and (\ref{MC+.s0-l.1}),
%to momentum conservation from the function $f_A(x)$,
it is convenient to calculate the following  auxiliary integral
\be
I(\mu)= \int_0^1 dx\, x^{\mu} \, f_A(x)= \int_0^1 dx \, \frac{x^{\mu}}{\ln \frac{1}{1-x} +a+ \Psi(\nu+2)}\, \frac{(1-x)^{\nu +1}}{\Gamma(\nu+2)}\,.
\label{Imu}
\ee

Expanding the denominator as in Eq. (\ref{Af.exp}),
%it was done above,
we have
\be
I(\mu)=  \sum_{m=0}^{\infty} \frac{(-1)^m}{a^{m +1}} \int_0^1 dx \, x^{\mu} \, \left(\ln \frac{1}{1-x} + \Psi(\nu+2)\right)^m\, \frac{(1-x)^{\nu +1}}{\Gamma(\nu+2)}
\label{Imu.1}
\ee

As it was shown above in Eqs. (\ref{Afx}) and (\ref{Afx.1}) the integral in the r.h.s. has the following form
\bea
&&\int_0^1 dx x^{\mu} \, \left(\ln \frac{1}{1-x} + \Psi(\nu+2)\right)^m\, \frac{(1-x)^{\nu +1}}{\Gamma(\nu+2)} \approx (-1)^m {\left(\frac{d}{d\nu}\right)}^m
\int_0^1 dx\, x^{\mu} \, \frac{(1-x)^{\nu +1}}{\Gamma(\nu+2)}\nonumber \\
&&=(-1)^m {\left(\frac{d}{d\nu}\right)}^m \, \frac{\Gamma(\mu +1)}{\Gamma(\mu+\nu+3)}
\approx \Psi^m(\mu+\nu+3) \, \frac{\Gamma(\mu +1)}{\Gamma(\mu+\nu+3)}\,.
\label{Int}
\eea

So, for the  auxiliary integral $I(\mu)$ we have
\be
I(\mu)\approx  \sum_{m=0}^{\infty} \frac{(-1)^m}{a^{m +1}} \,  \Psi^m(\mu+\nu+3) \, \frac{\Gamma(\mu +1)}{\Gamma(\mu+\nu+3)}= 
\frac{1}{a+ \Psi(\mu+\nu+3)}\, \frac{\Gamma(\mu +1)}{\Gamma(\mu+\nu+3)}
\label{Imu.2}
\ee

Thus, the results  $\sim K_{-}$ for $G_g^{-}(s)$ in Eq. (\ref{G+s}) and for $G_g^{-}(s=0)$ in Eq. (\ref{MC+.s0-l.1}),
can be obtained from (\ref{Imu.2}) for $\mu=1$.

To obtain the results (\ref{Phik}) it is convenient to consider 
the following  auxiliary integral
\be
\Phi_j(\mu,\nu)= \int_0^1 dx x^{\mu} \rho^j I_j (\sigma) \,(1-x)^{\nu}\,.  
\label{phimu}
\ee

Expanding Bessel function, we have
\be
\Phi_j(\mu,\nu)= \int_0^1 dx x^{\mu} \, \sum_{k=0}^{\infty} \, \frac {(ds)^{k+j}}{k!(k+j)!}\, \left(\ln \frac{1}{x}\right)^k (1-x)^{\nu} \,.  
\label{phimu.1}
\ee

As it was above in Eq. (\ref{Int}), the integral in the r.h.s. can be rewritten as
\be
\int_0^1 dx x^{\mu} \, \left(\ln \frac{1}{x}\right)^k (1-x)^{\nu} =(-1)^k {\left(\frac{d}{d\mu}\right)}^{k} \int_0^1 dx x^{\mu} \, (1-x)^{\nu}
= (-1)^k {\left(\frac{d}{d\mu}\right)}^{k} \, \frac{\Gamma(\mu+1)\Gamma(\nu+1)}{\Gamma(\mu+\nu+2)} \,.
\label{Inte}
\ee

Taking approximations (\ref{Psi}), we have
\be
\int_0^1 dx x^{\mu} \, \left(\ln \frac{1}{x}\right)^k (1-x)^{\nu} =
\bigl(\Psi(\mu+\nu+2)-\Psi(\mu+1)\bigr)^k  \, \frac{\Gamma(\mu+1)\Gamma(\nu+1)}{\Gamma(\mu+\nu+2)} \,.
\label{Inte.1}
\ee
and, thus, the integral $\Phi_j(\mu,\nu)$ is equal to
\be
\Phi_j(\mu,\nu) = \rho_{\mu}^j I_j (\sigma_{\mu}) \, \frac{\Gamma(\mu+1)\Gamma(\nu+1)}{\Gamma(\mu+\nu+2)} \,,
\label{phimu}
\ee
where
\be
\sigma_{\mu}=2\sqrt{ds(\Psi(\mu+\nu+2)-\Psi(\mu+1))},~~\rho_{\mu}=\frac{\sigma_{\mu}}{2(\Psi(\mu+\nu+2)-\Psi(\mu+1))}\,.
\label{simu}
\ee

The results (\ref{Phik}) for $\Phi_{j}(\nu)$ $(j=0,1)$ can be obtained from Eq. (\ref{phimu}) at $\mu=0$.

%\appendix
\def\theequation{C\arabic{equation}}
\setcounter{equation}{0}

\section{%TMD parton densities in
  Differential formulation of KMR approach}

%\subsection{TMDs from differential formulation of KMR approach} \noindent

Now we can
%use (\ref{Def}) to
find the results for TMD parton densities without derivatives.
Derivation of $T_a(Q^2,k^2)$ is as follows
\be
\frac{\partial T_a(Q^2,k^2)}{\partial \ln k^2} = d_a \, \beta_0 \, a_s(k^2) \, R_a(\Delta) \, T_a(Q^2,k^2),
\label{DefTa}
\ee
\noindent
and derivations of conventional PDFs are as follows
\bea
\z \frac{\partial f_{V}(x,k^2)}{\partial \ln k^2} =\beta_0 \, a_s(k^2) \Biggl\{r_{V}\ln(1-x)\, f_{V}(x,k^2) 
- \Biggl[d_{V}(1-\lambda_V) \,A_V\; e^{-d_{V}(1-\lambda) s_2} (1-x) \nonumber \\
  &&+\bigl[p_V+r_{V}\Psi(1+\nu_V(s_2))\bigr] \frac{B_{V}(s_2)\, x}{\Gamma(1+\nu_{V}(s_2))} + \stackrel{*}{D_V}(s_2) x(1-x)\Biggr]
  (1-x)^{\nu_{V}(s_2)} \Biggr\}
\, , \nonumber \\
%\label{DefV}\\
\z \frac{\partial f_{g,+}(x,k^2)}{\partial \ln k^2} =\beta_0 \, a_s(k^2) \Biggl\{ r_{+}\ln(1-x)\, f_{g,+}(x,k^2) 
-\Biggl[\left[
    %\frac{\hat{d}_+}{\rho_2} I_1(\sigma_2) + \overline d_{+} I_0(\sigma_2)
\hat{d}_+ \, \overline{I}_1(\sigma_2) + \overline d_{+} \overline{I}_0(\sigma_2)
    %+ d_{-} f_a^-(x,k^2)
  \right] \,A_g^{+}\; e^{-\overline d_{+} s_2} \nonumber \\
  \z \times  (1-x)^{m_{g+}} +\left[p_++r_{+}\Psi(1+\nu_+(s_2))\right] \frac{B_{+}(s_2)\, x}{\Gamma(1+\nu_{+}(s_2))} + \stackrel{*}{D^+_g}(s_2)  x(1-x)\Biggr]
  (1-x)^{\nu_{+}(s_2)}  \Biggr\}
\, ,  \nonumber \\
%\label{Deffg+}\\
\z \frac{\partial f_{q,+}(x,k^2)}{\partial \ln k^2} =\beta_0 \, a_s(k^2) \Biggl\{r_{+}\ln(1-x)\, f_{q,+}(x,k^2) \nonumber \\ 
\z
%-\beta_0 \, a_s(k^2)
\left[\hat{d}_+ \overline{I}_0(\sigma_2) + \overline d_{+} \,  \tilde{I}_1(\sigma_2)
%  + \overline d_{+} \rho_2  I_1(\sigma_2)
  %+ d_{-} f_a^-(x,k^2)
  \right] \,A_q^{+} \; e^{-\overline d_{+} s_2} (1-x)^{\nu_{+}(s_2)+1+m_{q+}} \Biggr\}\,, \nonumber \\
%\nonumber \\
%&&-\left[p_++r_{+}\Psi(1+\nu_+(s))\right] \frac{B_{+}(s)\, x}{\Gamma(1+\nu_{+}(s))} + ...
%\, ,
%\label{Deffq+}\\
%\z \frac{\partial f_{q,-}(x,k^2)}{\partial \ln k^2} =\beta_0 \, a_s(k^2) \Biggl\{ r_{-}\ln(1-x)\, f_{q,-}(x,k^2) 
%-\Biggl[d_{-} \,A_q\; e^{d_{-} s_2} (1-x)^{m_{q-}} \nonumber \\
%  \z+\left[p_-+r_{-}\Psi(1+\nu_-(s_2))\right] \frac{B_{-}(s_2)\, x}{\Gamma(1+\nu_{-}(s_2))}+ \stackrel{*}{D^-_q}(s_2)  x(1-x)\Biggr]
%  (1-x)^{\nu_{-}(s_2)}  \Biggr\}
%\, , \nonumber \\
%%\label{Deffq-}\\
%\z \frac{\partial f_{g,-}(x,k^2)}{\partial \ln k^2} =\beta_0 \, a_s(k^2) \Biggl\{  r_{-}\ln(1-x)\, f_{g,-}(x,k^2) 
%-\biggl[ d_{-} \,A^{-}_g\; e^{d_{-} s_2} (1-x)^{m_{g-}} \nonumber \\
%\z+%\left[p_-+r_{-}\Psi(2+\nu_-(s))\right]
%\frac{K_{-}\,B_-(s_2) x}{\Gamma(2+\nu_{-}(s_2))}  \,
%  \frac{p_-+r_{-}\Psi(2+\nu_-(s_2))}{\left[\ln(1/(1-x))+\hat{c} + \Psi(\nu_{-}(s_2)+2)\right]} \biggr]  
%  \,  (1-x)^{\nu_{-}(s_2)+1} \Biggr\}
\z \frac{\partial f_{q,-}(x,k^2)}{\partial \ln k^2} =\beta_0 \, a_s(k^2) \Biggl\{ r_{-}\ln(1-x)\, f_{q,-}(x,k^2) 
-\Biggl[d_{-} \,A_q\; e^{d_{-} s_2} (1-x)^{m_{q-}} \nonumber \\
  \z
  %+\left[p_-+r_{-}\Psi(1+\nu_-(s))\right] \frac{B_{-}(s)\, x}{\Gamma(1+\nu_{-}(s))}
  + \stackrel{*}{D^-_q}(s_2)  x(1-x)\Biggr]
  (1-x)^{\nu_{-}(s_2)}  \Biggr\}
\, , \nonumber \\
%\label{Deffq-.1}\\
\z \frac{\partial f_{g,-}(x,k^2)}{\partial \ln k^2} =\beta_0 \, a_s(k^2) \Biggl\{  r_{-}\ln(1-x)\, f_{g,-}(x,k^2) 
-
%\biggl[
d_{-} \,A^{-}_g\; e^{d_{-} s_2}
%(1-x)^{m_{g-}} \nonumber \\
%\z+%\left[p_-+r_{-}\Psi(2+\nu_-(s))\right]
%\frac{K_{-}\,B_-(s) x}{\Gamma(2+\nu_{-}(s))}  \,
 % \frac{p_-+r_{-}\Psi(2+\nu_-(s))}{\left[\ln(1/(1-x))+\hat{c} + \Psi(\nu_{-}(s)+2)\right]} \biggr]  
  \,  (1-x)^{\nu_{-}(s_2)+m_{g-}+1} \Biggr\}
\, ,
\label{Deffg-}
\eea
where
%the symbol $\stackrel{*}{D_{...}}(s)
%$...$
%marks contribtuions form the derivations of $D_{\pm}(s)$, which should be small.
\be
\stackrel{*}{D_{...}}(s)=\frac{d}{ds} \, D_{...}(s)\,,~~ s_2=\ln \left( \frac{a_s(Q^2_0)}{a_s(k^2)} \right),~~
\sigma_2=\sigma(s\to s_2),~~\rho_2=\rho(s\to s_2)\,.
\label{DerD}
\ee

So, the results for the TMD parton densities read the form (\ref{un_fpm}) with
%becomes to be
\bea
\z f^{(d)}_{V}(x,k^2,Q^2) = \beta_0 \, a_s(k^2) T_q(Q^2,k^2) \times \Biggl\{ \left[d_q \, R_q(\Delta)-r_{V} \ln\left(\frac{1}{1-x}\right)\right] 
    \, f_{V}(x,k^2) \nonumber \\
    \z - \Biggl[d_{V}(1-\lambda_V) \,A_V\; e^{-d_V(1-\lambda) s_2} (1-x) \nonumber \\
      &&+\left[p_V+r_{V}\Psi(1+\nu_V(s_2))\right] \frac{B_{V}(s_2)\, x}{\Gamma(1+\nu_{V}(s_2))}+ \stackrel{*}{D_v}(s_2) x(1-x)\Biggr]
      (1-x)^{\nu_{V}(s_2)} \Biggr\}
\, , \nonumber \\
%\label{UnfV}\\
\z f^{(d)}_{g,+}(x,k^2,Q^2) = \beta_0 \, a_s(k^2) T_g(Q^2,k^2) \times \Biggl\{ \left[d_g \, R_g(\Delta)-r_{+} \ln\left(\frac{1}{1-x}\right)\right] 
    \, f_{g,+}(x,k^2) \nonumber \\
    \z -\Biggl[\left[\hat{d}_+ \overline{I}_1(\sigma_2)+ \overline d_{+} \overline{I}_0(\sigma_2)
        %\frac{\hat{d}_+}{\rho_2} I_1(\sigma_2) + \overline d_{+} I_0(\sigma_2)
  %+ d_{-} f_a^-(x,k^2)
        \right] \,A_g^{+}\; e^{-\overline d_{+} s_2} (1-x)^{m_{g+}}
      \nonumber \\
      &&+\left[p_+ + r_{+}\Psi(1+\nu_+(s_2))\right] \frac{B_{+}(s_2)\, x}{\Gamma(1+\nu_{+}(s_2))}   + \stackrel{*}{D^+_g}(s_2)  x(1-x)\Biggr]
     (1-x)^{\nu_{+}(s_2)} \Biggr\}
\, , \nonumber \\
%\label{Unfg+}\\
  \z f^{(d)}_{q,+}(x,k^2,Q^2) = \beta_0 \, a_s(k^2) T_q(Q^2,k^2) \times \Biggl\{ \left[ d_q \, R_q(\Delta)-r_{+} \ln\left(\frac{1}{1-x}\right)\right] 
    \, f_{q,+}(x,k^2) \nonumber \\
    \z -\left[\hat{d}_+\, \overline{I}_0(\sigma_2) + \overline d_{+} \tilde{I}_1(\sigma_2)
      %\rho_2 I_1(\sigma_2)
  %+ d_{-} f_a^-(x,k^2)
  \right] \,A_q^{+}\; e^{-\overline d_{+} s_2} (1-x)^{\nu_{+}(s_2)+1+m_{q+}} \Biggr\}\, , \nonumber \\
%\label{Unfq+}\\
\z f^{(d)}_{q,-}(x,k^2,Q^2) = \beta_0 \, a_s(k^2) T_q(Q^2,k^2) \times \Biggl\{ \left[d_q \, R_q(\Delta)-r_{-} \ln\left(\frac{1}{1-x}\right)\right] 
    \, f_{q,-}(x,k^2) \nonumber \\
    \z -\Biggl[d_{-} \,A_q\; e^{-d_{-} s_2} (1-x)^{m_{q-}}
      %\nonumber \\
      %&&+\left[p_-+r_{-}\Psi(1+\nu_-(s_2))\right] \frac{B_{-}(s_2)\, x}{\Gamma(1+\nu_{-}(s_2))}
      + \stackrel{*}{D^-_q}(s_2)  x(1-x)\Biggr]
      (1-x)^{\nu_{-}(s_2)} \Biggr\}
\, , \nonumber \\
%\label{Unfq-}\\
\z f^{(d)}_{g,-}(x,k^2,Q^2) = \beta_0 \, a_s(k^2) T_g(Q^2,k^2) \times \Biggl\{ \left[d_g \, R_g(\Delta)-r_{-} \ln\left(\frac{1}{1-x}\right)\right] 
    \, f_{g,-}(x,k^2) \nonumber \\
    \z -
    %\Biggl[
    d_{-} \,A_g^{-}\; e^{-d_{-} s_2}
    %(1-x)^{m_{g-}}
    %\nonumber \\&&
    %+\left[p_-+r_{-}\Psi(1+\nu_-(s_2))\right] \nonumber \\&& \times
 %\frac{K_{-}}{\Gamma(2+\nu_{-}(s_2))}  \,
 % \frac{B_-(s_2) x}{\left[\ln(1/(1-x))+\hat{c} + \Psi(\nu_{-}(s_2)+2)\right]} \Biggr]
  \,  (1-x)^{\nu_{-}(s_2)+m_{g-}+1} \Biggr\}
\, .
\label{Unfg-}
\eea

\noindent
%The expressions for conventional PDFs and Sudakov form factors are given in the Sections~2 and~3.1, respectively.
%Since the value of $\hat{d}_+$ is negative and the factor $\hat{d}_+/\rho_a$ is large at low $x$, the TMDs $f^{(d)}_a(x,k^2,\mu^2)$ are
%positive at small $x$. With increasing $x$, the results
%for the latter can be negative.\\% that, in particular, demonstrates an inapplicability of(\ref{8.02}) for PDFs.

\end{document}